\documentclass[12pt, letterpaper]{article}

\RequirePackage[hyphens]{url}
\usepackage[utf8]{inputenc}

\usepackage[export]{adjustbox}
\usepackage{amsmath}
\usepackage{amssymb}
\usepackage{animate}
\usepackage{appendix}
\usepackage{array}
\usepackage{authblk}
\usepackage{bm}
\usepackage{booktabs}
\usepackage{caption}
\usepackage{color}
\usepackage{changepage}
\usepackage{chemformula}
\usepackage{csvsimple}
\usepackage{etoolbox}
\usepackage{enumitem}
\usepackage{float}
\usepackage{framed}
\usepackage[margin=1in]{geometry}
\usepackage{graphicx}
\definecolor{links}{HTML}{0074CC} 
\usepackage[colorlinks=true, allcolors=links]{hyperref}
\usepackage{lineno}
\usepackage{lscape}
\usepackage{listings}
\usepackage{makecell}
\usepackage{marginnote}
\usepackage{mathrsfs}
\usepackage{mathtools}
\usepackage{mdframed}
\usepackage{multicol}
\usepackage{multirow}
\usepackage[natbibapa]{apacite}
\usepackage{setspace}
\usepackage{soul}
\usepackage{tabulary}
\usepackage{subcaption}
\usepackage{tikz}
\usetikzlibrary{shapes, arrows, positioning, fit}
\usepackage{wrapfig}
\usepackage{xargs}
\usepackage[normalem]{ulem}
\graphicspath{ {figs/} }
\usepackage{epstopdf}
\AppendGraphicsExtensions{.tif}
\PassOptionsToPackage{hyphens}{url}\usepackage{hyperref}

\definecolor{sandstorm}{rgb}{0.93, 0.84, 0.25}

 \bibpunct[, ]{(}{)}{,}{a}{}{,}%
\newcommand{\instructions}[1]{}

\newcolumntype{C}[1]{>{\centering\arraybackslash}p{#1}}

\providecommand{\keywords}[1]
{
  \noindent \small	
  \textbf{Keywords: } #1
}

\makeatletter
\newcommand*{\centerfloat}{%
  \parindent \z@
  \leftskip \z@ \@plus 1fil \@minus \textwidth
  \rightskip\leftskip
  \parfillskip \z@skip}
\makeatother

\AtBeginEnvironment{tabular}{\footnotesize} 
\usepackage[colorinlistoftodos,prependcaption,textsize=tiny]{todonotes}
\newcommandx{\unsure}[2][1=]{\todo[linecolor=red,backgroundcolor=red!25,bordercolor=red,#1]{#2}}
\newcommandx{\change}[2][1=]{\todo[linecolor=blue,backgroundcolor=blue!25,bordercolor=blue,#1]{#2}}
\newcommandx{\info}[2][1=]{\todo[linecolor=orange,backgroundcolor=orange!25,bordercolor=orange,#1]{#2}}
\newcommandx{\improvement}[2][1=]{\todo[linecolor=sandstorm,backgroundcolor=	sandstorm!25,bordercolor=sandstorm,#1]{#2}}
\newcommandx{\thiswillnotshow}[2][1=]{\todo[disable,#1]{#2}}

\definecolor{miamired}{HTML}{C3142D}

\newif\ifblinded
\newif\iffinal

\newcommand{\setfinal}{
    \blindedfalse
    \finaltrue
}

\newcommand{\authorinfo}{
    \ifblinded
        \author{\large (Authors blinded for peer review)}
    \fi
    \iffinal
        \author[1,*]{\"Ozge S\"urer}
        \affil[1]{Farmer School of Business, Miami University, Oxford, OH 45056, USA}
        \affil[*]{Corresponding author can be reached at surero@miamioh.edu}
    \fi
}

\def\hb#1{[BLINDED FOR REVIEW]}

\title{\Large \textbf{Simulation Experiment Design for Calibration via Active Learning}}

\usepackage{caption}
\usepackage{subcaption}
\usepackage{graphicx}
\usepackage{bbm}
\usepackage{times,scalefnt}
\usepackage{amsmath,amssymb,mathtools}
\usepackage{tikz}
\usepackage{diagbox}
\DeclareMathOperator*{\argmax}{arg\,max}
\DeclareMathOperator*{\argmin}{arg\,min}

\usepackage[boxruled,vlined,linesnumbered,commentsnumbered]{algorithm2e}
\usepackage{subcaption}

\newcommand{\p}{{p}}

\newcommand{\xb}{{\mathbf{x}}} 
\newcommand{\zb}{{\mathbf{z}}} 

\newcommand{\y}{{\mathbf{y}}}
\newcommand{\model}{{\pmb{\eta}}}

\newcommand{\thetav}{{\pmb{\theta}}}
\newcommand{\Sigmav}{{\pmb{\Sigma}}}

\newcommand{\Kv}{{\mathbf{K}}}
\newcommand{\kv}{{\mathbf{k}}}

\newcommand{\Sv}{{\mathbf{S}}}

\newcommand{\phiv}{{\pmb{\phi}}}
\newcommand{\muv}{{\pmb{\mu}}}
\newcommand{\rhov}{{\pmb{\zeta}}}
\newcommand{\PHI}{\phiv_{t}\left(\thetav, \zb^*\right)}
\newcommand{\cov}{\Sv_t\left(\pmb{\theta}\right)}
\newcommand{\meanv}{\pmb{\mu}_t\left(\pmb{\theta}\right)}
\newcommand{\tcb}{\textcolor{black}}

\usepackage[english]{babel}
\usepackage{amsthm}
\newtheorem{theorem}{Theorem}[section]

\newtheorem{lemma}[theorem]{Lemma}

\setfinal

\date{\small \today}

\begin{document}

\authorinfo

\maketitle

\begin{abstract}
\noindent Simulation models often have parameters as input and return outputs to understand the behavior of complex systems.
    Calibration is the process of estimating the values of the parameters in a simulation model in light of observed data from the system that is being simulated. 
    When simulation models are expensive, emulators are built with simulation data as a computationally efficient approximation of an expensive model. An emulator then can be used to predict model outputs, instead of repeatedly running an expensive simulation model during the calibration process.
    Sequential design with an intelligent selection criterion can guide the process of collecting simulation data to build an emulator, making the calibration process more efficient and effective.
    This article proposes two novel criteria for sequentially acquiring new simulation data in an active learning setting by considering uncertainties on the posterior density of parameters. 
    Analysis of several simulation experiments and real-data simulation experiments from epidemiology demonstrates that proposed approaches result in improved posterior and field predictions.

\end{abstract}

\keywords{acquisition, Bayesian calibration, emulation, sequential design, uncertainty quantification}





 


\thispagestyle{empty}

\newpage 

\doublespacing

\section{Introduction}
\label{sec:intro}

Simulation models are pervasive in many engineering and science disciplines to explain the behavior of complex systems. Examples include the simulation of inventory/supply chain systems \citep{Hong2006}, manufacturing systems \citep{Chen2018}, storm surge \citep{Plumlee2021}, and physics \citep{Surer2022}. In these and many other cases, in addition to controllable inputs (aka design inputs), simulation models take user-specified calibration parameters as input and return outputs that can be used to understand reality. However, these parameters are often unknown and need to be inferred using observed data from a real physical/field experiment. Calibration is a way to infer parameters to ensure that simulation outputs accurately represent the real-world system that is being simulated. Calibration becomes a more challenging problem when a single simulation evaluation requires a significant amount of computational resources and time. Therefore, careful selection of the simulation experiments to run is a critical concern. 

This work considers Bayesian calibration, which is a specialized form of calibration that offers a way to quantify uncertainty in both parameters and predictions of quantities of interest. In a standard Bayesian calibration, emulators are built via statistical models such as Gaussian processes  \citep{Rasmussen2005, gramacy2020surrogates} or Bayesian tree techniques \citep{Chipman1998, Gramacy2008} to mimic the behaviors of the computationally expensive simulation model. Emulators then provide predictions of simulation outputs at any input configuration without running the expensive simulation model. Emulators are built with a simulation data set consisting of a set of inputs (called design) and corresponding simulation outputs. Once an emulator is constructed, it is used to facilitate the calibration process \citep{Ohagan2001, Higdon2004}. Hence, the precision of the calibration process relies on the emulator's accuracy. Common techniques to generate designs include random sampling and space-filling designs such as Latin hypercube sampling (LHS, \cite{McKay1979}), minimax designs \citep{Johnson1990} and the optimized versions \citep{Joseph2015}; see \cite{santner2018design} for a detailed survey. One drawback of such designs is that the (unknown) input region of interest may not be adequately explored, especially in high-dimensional spaces, leading to potential gaps in the emulator's predictive ability. Simulation experiment design, the topic of this article, should be selected with care to achieve precise calibration inference with a limited number of simulation runs.

Sequential design or active learning allows adaptive simulation data collection based on the simulation data set that has already been gathered. In a sequential design, the decision to sample additional data points is often based on statistical criteria, called acquisition function. The adaptability of the sequential design allows practitioners to focus on regions of interest or refine the experiment as it progresses. Moreover, sequential designs are often more resource-efficient than one-shot design procedures mentioned above since the iterative data collection can be terminated when a sufficient level of preciseness is achieved \citep{Lam2008}. Bayesian optimization (BO) \citep{Frazier2018} is a common sequential approach to optimize a black-box, computationally expensive function. In BO, an emulator is used to approximate the unknown objective function and then an acquisition function guides the selection of a new point to evaluate the objective function next. Expected improvement and probability of improvement are commonly used acquisition functions in BO to select the input with the highest expectation/probability for improvement over the best objective value obtained thus far \citep{Jones1998}.

Our approach stands out due to its distinctive utilization of the emulation strategy employed in modern calibration techniques unlike many existing methods in the literature on sequential design, which often do not directly emulate the simulation model itself but rather focus on emulating objective functions (i.e., goodness-of-fit measures). For instance, in the calibration context, \cite{Kandasamy2015} build an emulator of the log-likelihood to estimate the posterior density of parameters. \cite{Joseph2015b} and \cite{Joseph2019} introduce an energy design criterion to obtain a sample from the probability density function using an emulator of the density itself. Constructing emulators for objective functions can pose challenges compared to directly developing emulators of simulation models. One downside of direct emulation of the objective function is that the internal structure of the complex simulation models is often missed. Moreover, transferring the simulation output to obtain the objective function value brings extra complexity to the inference problem. Successful examples of integrating emulators of simulation models within a sequential framework can be found in \cite{Damblin2018, Koermer2023, Surer2023, Lartaud2024}. This work constructs emulators as proxies for simulation models to make the calibration process more efficient and effective by leveraging the information hidden in the structure of the models.

Sequential design has been used in the literature to build globally accurate emulators of simulation models. For a global emulator, one natural acquisition function is to choose the next input with the highest emulation variance \citep{Sacks1989, Seo200}. In parallel to this, \cite{MacKay1992} employs an active learning setting where inputs are selected based on the entropy criterion. This approach demonstrates that selecting inputs with the highest emulation variance approximates a maximum entropy design. However, since accurate predictions are desired across the entire input space, a criterion relying on a single point's uncertainty often leads to suboptimal results. The integrated mean squared prediction error (IMSPE), which considers the aggregated emulator uncertainty across the input space, is one of the most common acquisition functions in this field; for a more thorough review see \cite{gramacy2020surrogates}. Since it is theoretically sound and applicable in practice, there are numerous developments of IMSPE; see examples in  \cite{Binois2018, Cole2021, Sauer2022}. Because IMSPE for a general-purpose emulator does not take into account the observed data, it does not bring any additional advantage for calibration inference. Recently, \cite{Koermer2023} propose a novel IMSPE criterion within the Kennedy and O’Hagan calibration framework to improve predictions, and \cite{Lartaud2024} develop a weighted IMSPE criterion for Bayesian inverse problems. In this work, our focus is on the aggregated uncertainty in the estimate of the posterior density of parameters rather than the emulator uncertainty.

In a recent work, \cite{Surer2023} propose the expected integrated variance (EIVAR) criterion to accurately learn the posterior density of parameters. In their setting, when run at a parameter, the simulation model returns a high-dimensional output consisting of multiple responses collected on a set of fixed design inputs. Consequently, the only simulation inputs involved are parameters, and EIVAR is derived to sequentially acquire a new parameter and its high-dimensional output. However, in many settings, a simulation model is often a function of both a parameter and a design input (see the problem relating to a diaper line from the Procter \& Gamble Company in \cite{Krishna2022}, examples from nuclear physics in \cite{Phillips2021}, and the industrial application involving a chemical process in \cite{Koermer2023}). A design input is shared by both the simulation model and the field experiment. At a set of design inputs, which we call field data design inputs throughout the paper, field experiments are conducted to explore a physical system, and field data measured from the experiments are used to infer the unknown parameters of the simulation model. In this work, we first derive the EIVAR criterion to allow the acquisition of a simulation input consisting of a parameter and a design input simultaneously. The acquisition function encourages the selection of simulation design inputs aligned with field data design inputs around the parameter region of interest, and we demonstrate that such acquisitions lead to improved posterior predictions. However, when the calibration experiments target field predictions at unseen design inputs as well, focusing only on the field data design inputs does not allow exploration of the entire design space. For improved field predictions under novel design inputs, we propose another acquisition function by considering the uncertainties in the posterior obtained with unseen design inputs. The acquisition function prefers matching the simulation design inputs with the field data design inputs as well as exploring the remaining design space. Similar to our findings, \cite{Ranjan2011} suggest the alignment of the simulation design inputs with the field data design inputs for the field prediction. However, only a one-step simulation data collection is empirically conducted in \cite{Ranjan2011} and a systematic way is not proposed to acquire a new input. Our sequential approach can be considered an automated way to find how to allocate the simulation data to both existing and unseen design inputs for improved calibration inference.

The remainder of the paper is organized as follows. 
Section~\ref{sec:review} presents the main steps of the sequential design procedure as well as background on Bayesian calibration and Gaussian processes. Section~\ref{sec:acquisition} contains our methodologic contributions.  
In Section~\ref{sec:experiment}, we demonstrate the benefits of our proposed methods by analyzing results from several simulation experiments including synthetic models and a COVID-19 simulation model. Conclusions are presented in Section~\ref{sec:conc}.

\section{Background}
\label{sec:review}
This section overviews the proposed sequential algorithm, Bayesian calibration, and Gaussian process emulators.

\subsection{Sequential Experimental Design}
\label{sec:overview}

We use the following notations throughout the paper. Boldface characters are used to represent vectors and matrices. We label the length-$q$ vector of design inputs $\xb$ and the length-$p$ vector of calibration parameters $\thetav$. The simulation model, represented with $\eta(\cdot, \cdot)$, is a function that takes design input $\xb$ in space $\mathcal{X} \subset \mathbb{R}^{q}$ and parameter $\thetav$ in space $\Theta \subset \mathbb{R}^{p}$ as an input and returns one-dimensional simulation output $\eta\left(\xb, \thetav\right) \in \mathbb{R}$. Our goal is to sequentially collect simulation outputs from $n$ acquired inputs for improved calibration inference.

\begin{algorithm}[H]  
    \textbf{Input:} An initial sample size $n_0$, a total number of acquisitions $n$, a simulation model $\eta(\cdot, \cdot)$, and an acquisition function $\mathcal{A}_t(\cdot, \cdot)$

    \emph{Initialize} $\mathcal{D}_1 = \{((\xb_i, \thetav_{i}), \eta(\xb_i, \thetav_{i})) : i = 1, \ldots, n_0\}$
    
    \For {$t = 1,\ldots,n$} {
        \emph{Fit} an emulator with $\mathcal{D}_{t}$
        
        \emph{Generate} candidate solutions $\mathcal{L}_t$
        
        \emph{Select} $(\xb^{\rm new}, \thetav^{\rm new}) \in \argmin\limits_{(\xb^*, \thetav^*) \in \mathcal{L}_t} \mathcal{A}_t(\xb^*, \thetav^*)$

        \emph{Evaluate} $\eta(\xb^{\rm new}, \thetav^{\rm new})$

        \emph{Update} 
        $\mathcal{D}_{t+1} \gets \mathcal{D}_{t} \cup ((\xb^{\rm new}, \thetav^{\rm new}), \eta(\xb^{\rm new}, \thetav^{\rm new}))$
        }

    \textbf{Output:} Simulation data $\mathcal{D}_{n+1}$ along with the emulator fitted with $\mathcal{D}_{n+1}$
    
    \caption{Sequential experimental design}
    \label{alg:oaat}
\end{algorithm}

The proposed sequential design is summarized in Algorithm~\ref{alg:oaat}. The algorithm starts with an initial set of $n_0$ simulation inputs and their outputs stored in $\mathcal{D}_1 = \{((\xb_i, \thetav_{i}), \eta(\xb_i, \thetav_{i})) : i = 1, \ldots, n_0\}$ (see line~2). The initial simulation data set is typically sampled randomly from a prior distribution or through a space-filling design. During each iteration indexed by $t$, a new input is acquired and the simulation model is evaluated with the new input. The simulation data set $\mathcal{D}_{t + 1} = \{((\xb_{i}, \thetav_{i}), \eta(\xb_{i}, \thetav_{i})): i = 1, \ldots, n_t\}$, where $n_t = n_0 + t$, keeps all the simulation data obtained by the end of iteration $t$ (line~8). At the beginning of each iteration, an emulator is fitted to the simulation data set $\mathcal{D}_{t}$ (see Section~\ref{sec:GP}) and it is used to construct the acquisition function $\mathcal{A}_t(\cdot, \cdot)$ (see Section~\ref{sec:acquisition}). To avoid difficult numerical optimization, the acquisition function $\mathcal{A}_t(\cdot, \cdot)$ is minimized over a discrete set of inputs $\mathcal{L}_t$ to determine the next best input to evaluate the simulation model (see lines~5--7). Although the termination criteria for a sequential design can vary depending on the calibration objective, we use a fixed budget of $n$ acquired simulation outputs for comparison purposes.

\subsection{Bayesian Calibration}
\label{sec:notation}

The purpose of Bayesian calibration is to infer unknown calibration parameters using data from the field experiment and to characterize uncertainties in inferred parameters and associated predictions. Let $\xb^f_i$ be the design input where the field experiment is conducted and $y\left(\xb^f_i\right)$ denote the observed data from the field experiment for $i = 1, \ldots, d$. We first model the data from the field experiment
\begin{equation}
    y\left(\xb^f_i\right) = \eta\left(\xb^f_i, \thetav\right) + \epsilon,   \label{eq:statmodel}
\end{equation}
where $\epsilon \sim {\rm N}\left(0, \sigma^2\right)$ denotes the residual error. Let $\Sigmav$ be a $ d \times d$ diagonal error covariance matrix with diagonal elements $\sigma^2$. We assume $\sigma^2$ is known to account for the uncertainty in the difference between the data and the model and extend our criterion for the case of unknown $\sigma^2$ with a discrepancy term in Section~\ref{sec:acquisition}. Our results are built upon the assumption that the observation noise $\epsilon$ follows a normal distribution; we leave the derivations of proposed acquisition functions for different distributions of the residual error as future development.

In the Bayesian calibration framework, we are interested in the quantity $\p\left(\thetav|\y\right)$, which is the posterior probability density of parameter $\thetav$ given field data $\y = \left(y\left(\xb^f_1\right), \ldots, y\left(\xb^f_{d}\right)\right)^\top$. 
The initial knowledge about parameter $\thetav$ is represented by the prior probability density $p\left(\thetav\right)$, which is typically a known, closed-form function of the parameter.
Based on Bayes' rule, the posterior density has the form 
\begin{equation} 
\label{eq:posterior}
    p\left(\thetav|\y\right) = \frac{ p\left(\y|\thetav\right) p\left(\thetav\right)}{\int_{\Theta}  p\left(\y|\thetav'\right) p\left(\thetav'\right) {\rm d \thetav'}} \propto \tilde{p}\left(\thetav|\y\right) =  p\left(\y|\thetav\right) p\left(\thetav\right),
\end{equation}
where $p\left(\y|\thetav\right)$ is the likelihood function indicating the agreement between the simulation output at $\thetav$ and field data $\y$ and $\tilde{p}\left(\thetav|\y\right)$ represents the unnormalized posterior. In a typical Bayesian calibration, Markov chain Monte Carlo (MCMC) methods \citep{Gelman2004} are employed to produce samples from the posterior. Since the posterior is analytically intractable in complex models, the unnormalized posterior is used to represent the posterior up to a constant multiplier within MCMC schemes. This work considers the uncertainty in the estimate of the unnormalized posterior $\tilde{p}(\thetav|\y)$, which is referred to as the posterior for brevity throughout the remainder of this paper. 
The differences between the field data and the simulation outputs are assumed to follow a multivariate normal (MVN) distribution due to \eqref{eq:statmodel}, and the likelihood satisfies
\begin{equation}
   \p\left(\y|\thetav\right) = (2 \pi)^{-d/2} |\Sigmav|^{-1/2} \exp\left(-\frac{1}{2} \left(\y - \model\left(\thetav\right)\right)^\top \Sigmav^{-1} \left(\y - \model\left(\thetav\right)\right) \right), \label{eq:truelike}
\end{equation}
in which requires the simulation output $\model(\thetav) = \left(\eta\left(\xb^f_1, \thetav\right), \ldots, \eta\left(\xb^f_{d}, \thetav\right)\right)^\top$ at field data design inputs. To produce posterior samples, MCMC techniques have to evaluate $\p\left(\y|\thetav\right)$ many times (usually thousands or millions of evaluations) for candidate values of $\thetav$. However, the evaluation of a simulation model with any candidate parameter becomes computationally prohibitive when a single simulation run takes a substantial amount of computational time. To solve this problem, the emulator of a simulation model introduced in Section~\ref{sec:GP} can be used to estimate the posterior at any $\thetav$.

\subsection{Gaussian Process Model}
\label{sec:GP}
We consider Gaussian process (GP) modeling to build an emulator of the simulation model at each iteration of the sequential procedure based on the simulation data set $\mathcal{D}_t$ as shown in line~4 of Algorithm~\ref{alg:oaat}. GP emulators are commonly used for calibrating simulation models since GPs can provide both a predictive mean and variance for quantifying uncertainties. Our contribution involves the integration of the emulator with two novel acquisition functions for improved calibration inference via active learning (see Section~\ref{sec:acquisition}). For simplicity, we assume a zero-mean GP prior with the covariance defined by a positive definite kernel function $k_t(\cdot, \cdot) = \tau_t^{2} c(\cdot, \cdot; \rhov_t)$, a scaling parameter $\tau_t^{2}$ and a lengthscale parameter $\rhov_t = (\zeta_{t,1}, \ldots, \zeta_{t,{q+p}})^\top$. A scaling parameter $\tau_t^{2}$ controls the magnitude of the range of the simulation output represented by the GP to capture variations in the data, whereas a lengthscale parameter $\rhov_t$ controls the smoothness of the output (see Chapter~5.2 of \cite{gramacy2020surrogates} for a detailed survey of GP hyperparameters). The covariance can be parameterized by many different choices of kernel functions such as Gaussian and Mat\'ern \citep{Rasmussen2005, santner2018design}. The Gaussian kernel function is one of the most popular kernel functions due to its flexibility and theoretical properties and it typically works well for interpolating smooth functions. However, it is argued that strong smoothness assumptions might be unrealistic for many physical processes. Mat\'ern class of kernel functions is recommended to capture the variability of the underlying function better; see Chapter~4 of \cite{Rasmussen2005} for a detailed survey on different kernel functions. In this work, we use the separable version of the Mat\'ern correlation function with smoothness parameter 1.5 such that
\begin{equation}
    \begin{aligned}
        c(\mathbf{z}, \mathbf{z}'; \rhov_t) = & \prod_{l=1}^{q + p}\left[ (1 + |(z_l - z_l')\exp(\zeta_{t,l})|)  \exp\left(-\exp(\zeta_{t,l})|z_l - z_l'|\right)\right],
    \end{aligned}
\end{equation} 
where $\mathbf{z} = (z_1, \ldots, z_{q+p})^\top = \left(\xb^\top, \thetav^\top\right)^\top$ denotes a vector of size $q+p$ simulation input for notational simplicity. The choice of correlation function does not impact the rationale of the proposed acquisition functions.

Let the $n_t \times (q + p)$ matrix $\mathbf{z}_{1:n_t} = \left(\mathbf{z}_1, \ldots, \mathbf{z}_{n_t}\right)^\top$ represent the inputs where the simulation model has been evaluated. The simulation outputs are stored in $\model_{t} = \left(\eta(\xb_{1}, \thetav_{1}), \ldots, \eta(\xb_{n_t}, \thetav_{n_t})\right)^\top$. According to the GP prior, the joint distribution of the simulation outputs $\model_{t}$ and the output $\eta(\xb, \thetav)$ at an unseen input $\mathbf{z} = \left(\xb^\top, \thetav^\top\right)^\top$ is MVN distribution such that
\begin{equation}\label{gp:mvn}
    \begin{aligned}
        \left[{\begin{array}{c} \model_{t} \\
            \eta(\xb, \thetav) \\ \end{array}} \right]  \sim \text{MVN}\left( \mathbf{0},  \left[ {\begin{array}{cc} \Kv_t & \kv_t(\mathbf{z}) \\
                \kv_t(\mathbf{z})^\top & k_t(\mathbf{z}, \mathbf{z}) \\
                \end{array} } \right] \right).
    \end{aligned}
\end{equation}
Here, $\kv_t(\mathbf{z}) = \left(k_t\left(\mathbf{z}, \mathbf{z}_1\right), \ldots, k_t\left(\mathbf{z}, \mathbf{z}_{n_t}\right)\right)^\top$ is comprised of cross-kernel evaluations between $\mathbf{z}$ and $\mathbf{z}_{1:n_t}$ and $\Kv_t$ is the $n_t \times n_t$ matrix with $ij$ coordinates $k_t(\mathbf{z}_i,\mathbf{z}_j) + \upsilon \delta_{i=j}$ for $1 \leq i, j \leq n_t$. In addition, $\upsilon > 0$ is a nugget parameter and $\delta_{i=j}$ is the Kronecker delta function with value 1 if $i = j$ and value 0 otherwise. The nugget parameter $\upsilon$ is added to the diagonal elements of $\Kv_t$ to ensure the positive definiteness of the resulting matrix in the case of two identical inputs $\zb_i = \zb_j$ for improved numerical stability. Conditioning the joint GP prior distribution on $\model_{t}$ (see appendix A.2 of \cite{Rasmussen2005} for further details) reveals that the predictive distribution $\eta(\xb, \thetav)|\model_{t}$ is Gaussian with mean $m_{t}(\mathbf{z})$ and variance $\varsigma^2_{t}(\mathbf{z})$ such that $\eta(\xb, \thetav)|\model_{t} \sim \text{N}\left(m_{t}(\mathbf{z}), \varsigma^2_{t}(\mathbf{z})\right)$ where
\begin{equation}
    \begin{aligned}
         m_{t}(\mathbf{z}) = \kv_t(\mathbf{z})^\top \Kv_t^{-1} \model_{t} \text{ and }
         \varsigma^2_{t}(\mathbf{z}) = k_t(\mathbf{z}, \mathbf{z}) - \kv_t(\mathbf{z})^\top \Kv_t^{-1} \kv_t(\mathbf{z}). \label{eq:meanvar_latent} 
    \end{aligned}
\end{equation}
The emulator provides a probabilistic representation of the simulation output at design inputs with mean $\muv_t(\cdot)$ and covariance matrix $\Sv_t(\cdot)$ such that
\begin{equation}
    \model(\thetav)|\mathcal{D}_{t} \sim \text{MVN}\left(\meanv, \cov\right),
    \label{emu_final}
\end{equation}
where $\meanv = \left(m_{t}\left(\zb_1^f\right), \ldots, m_{t}\left(\zb_d^f\right)\right)^\top$ and the $i$th diagonal element of $\cov$ is $\varsigma^2_{t}\left(\zb_i^f\right)$ and $(i,j)$th element of $\cov$ is $\text{cov}_t\left(\zb_i^f, \zb_j^f\right) = k_t\left(\zb_i^f, \zb_j^f\right) - \kv_t\left(\zb_i^f\right)^\top \Kv_t^{-1} \kv_t\left(\zb_j^f\right)$ where $\zb_i^f = \left({\xb_i^f}^{\top}, \thetav^\top\right)^\top$ for $i, j = 1, \ldots, d$. The next section uses these results for posterior inference and the derivation of acquisition functions.

\section{Acquisition Functions}
\label{sec:acquisition}

The acquisition function $\mathcal{A}_t(\cdot, \cdot)$ in Algorithm~\ref{alg:oaat} provides a way to make an informed decision about where to evaluate the simulation model next given data $\mathcal{D}_t$. 
We propose two acquisition functions, one of which results in improved posterior predictions, while the other provides improved field predictions. 
The proposed acquisition functions focus on minimizing the aggregated variance of the posterior and use the mean $\mathbb{E}\left[\tilde{p}\left(\thetav|\y\right) | \mathcal{D}_{t}\right]$ and variance $\mathbb{V}\left[\tilde{p}(\thetav|\y) | \mathcal{D}_{t}\right]$ of posterior prediction from Lemma~\ref{lemma:UQ}. The proof follows from \cite{Surer2023} and is given in Appendix~\ref{app:3.1} for the sake of completeness. 
    \begin{lemma}\label{lemma:UQ}
    Assuming that the covariance matrices $\Sigmav$ and $\Sv_t(\thetav)$ are positive definite, under the model given by Equations~\eqref{eq:posterior}, ~\eqref{eq:truelike}, and ~\eqref{emu_final}, 
        \begin{gather}
            \mathbb{E}[\tilde{p}(\thetav|\y)| \mathcal{D}_{t}] = f_\mathcal{N}\left(\y; \, \meanv, \, \Sigmav + \cov\right) p(\thetav), \label{expectedpostfinal}\\
            \mathbb{V}[\tilde{p}(\thetav|\y) |\mathcal{D}_{t}] = \left(\frac{1}{2^{d}\pi^{d/2}|\Sigmav|^{1/2}}f_\mathcal{N}\left(\y; \, \meanv, \, \frac{1}{2}\Sigmav + \cov\right) \right. \nonumber \\
            \left. \hspace{2in} - \left(f_\mathcal{N}\left(\y; \, \meanv, \, \Sigmav + \cov\right)\right)^2\right)p(\thetav)^2, \label{variancepostfinal}
        \end{gather}
    where $f_\mathcal{N}(\mathbf{a}; \, \mathbf{b}, \, \mathbf{C})$ denotes the probability density function of the normal distribution with mean $\mathbf{b}$ and covariance $\mathbf{C}$, evaluated at the value $\mathbf{a}$.
    \end{lemma}
    
Recently, \cite{Surer2023} propose an expected integrated variance (EIVAR) criterion to select new simulation runs for an improved posterior prediction in the case of high-dimensional simulation outputs. Their setting considers that once the simulation model is evaluated with parameter $\thetav$, it returns simulation outputs simultaneously at all design inputs $\xb_1^f, \ldots, \xb_d^f$. In other words, when a new parameter is acquired at iteration $t$, the associated simulation outputs at all design inputs are included in the simulation data set. The GP-based emulator relying on the basis vector approach \citep{Higdon2008} is used to emulate the high-dimensional simulation output, and EIVAR is derived using this specific form of an emulator. First, we generalize the EIVAR criterion for a one-dimensional simulation output setting to allow the acquisition of both a design input and a parameter. The proposed acquisition function, denoted by $\mathcal{A}^{p}_t(\cdot)$, aggregates the variance of the posterior over the parameter space to better learn the posterior and is calculated for any candidate input $\mathbf{z}^* = \left({\xb^*}^{\top}, {\thetav^*}^{\top}\right)^\top$ from the discrete set of inputs $\mathcal{L}_t$ introduced in Section~\ref{sec:overview} by
\begin{align} \label{EIVAR_post}
    \begin{split}
        \mathcal{A}^{p}_t(\mathbf{z}^*) &= \int_{\Theta} \mathbb{E}_{\eta^* | \mathcal{D}_{t}} \left( \mathbb{V}[p(\y|\thetav) \left| (\mathbf{z}^*, \eta^*) \cup \mathcal{D}_{t} \right] \right) p(\thetav)^2 d \thetav.
    \end{split}
\end{align}
Here, $\eta^* \coloneqq \eta\left(\xb^*, \thetav^*\right)$ represents the new simulation output at $\mathbf{z}^*$ and the expectation is taken over the hypothetical simulation output $\eta^*$ under data $\mathcal{D}_t$. 

If accuracy over the entire design space $\mathcal{X}$ is desired then we suggest the following acquisition function represented by $\mathcal{A}^{y}_t(\cdot)$ for improved field predictions
\begin{align} \label{EIVAR_pred}
    \begin{split}
        \mathcal{A}^{y}_t(\mathbf{z}^*) &= \int_{\mathcal{X}} \mathbb{E}_{\eta^* | \mathcal{D}_{t}} \left( \mathbb{V}[p(\mathbf{y}^\xb|\hat{\thetav}) \left| (\mathbf{z}^*, \eta^*) \cup \mathcal{D}_{t} \right] \right) p(\hat{\thetav})^2 d \xb,
    \end{split}
\end{align}
where $\mathbf{y}^\xb = \left(y\left(\xb^f_1\right), \ldots, y\left(\xb^f_{d}\right), y(\xb)\right)^\top$ and $\hat{\thetav}$ represents the estimate of the parameter of interest. $\mathcal{A}^{y}_t(\cdot)$ integrates the variance of the posterior over the design space $\mathcal{X}$ by considering all design inputs as possible locations to collect field data for a given parameter estimate. Since the field data is available only at design inputs $\xb_1^f, \ldots, \xb_d^f$, we use the approximation of \eqref{EIVAR_pred} leveraging the predictions of the emulator to estimate $y(\xb)$ at any $\xb$ as described later in this section. Our focus here is on how to select the new input given the estimate. One can use the same plug-in estimate throughout the sequential procedure or obtain it at each iteration, both of which are illustrated in our experiments. Instead of substituting a single value $\hat{\thetav}$ into \eqref{EIVAR_pred}, another way is to integrate over the parameter space $\Theta$ as well to consider the total uncertainty on the posterior estimate across all parameters and design inputs. However, due to the computational cost accompanying the evaluation of double integral, we use the plug-in approach in our experiments and leave the computational enhancements as future work. 

\begin{figure}[t]
    \centering
    \begin{subfigure}{0.32\textwidth}
        \includegraphics[width=1\textwidth]{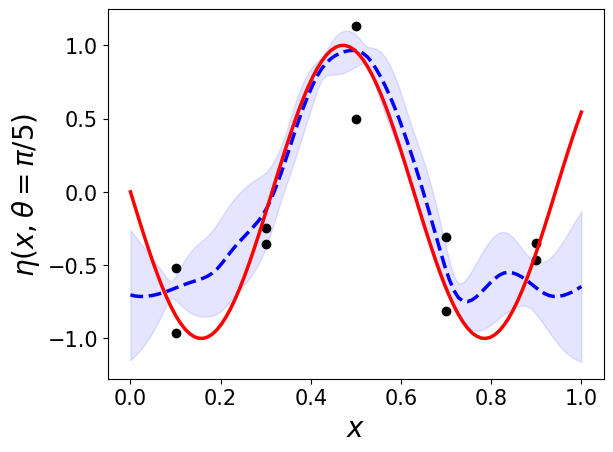}
    \end{subfigure}
    \begin{subfigure}{0.32\textwidth}
        \includegraphics[width=1\textwidth]{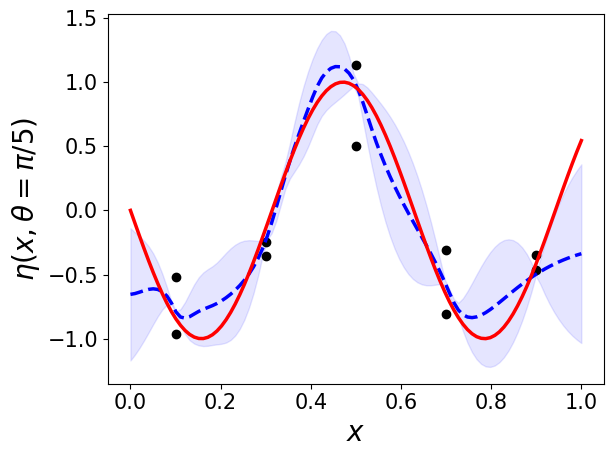}
    \end{subfigure}
    \begin{subfigure}{0.32\textwidth}
        \includegraphics[width=1\textwidth]{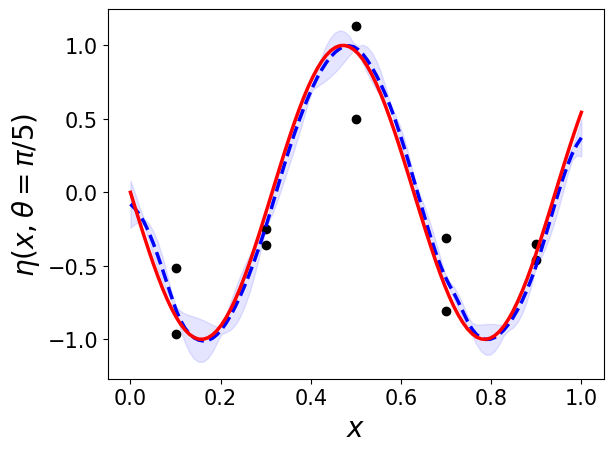}
    \end{subfigure}
    \begin{subfigure}{0.32\textwidth}
        \includegraphics[width=1\textwidth]{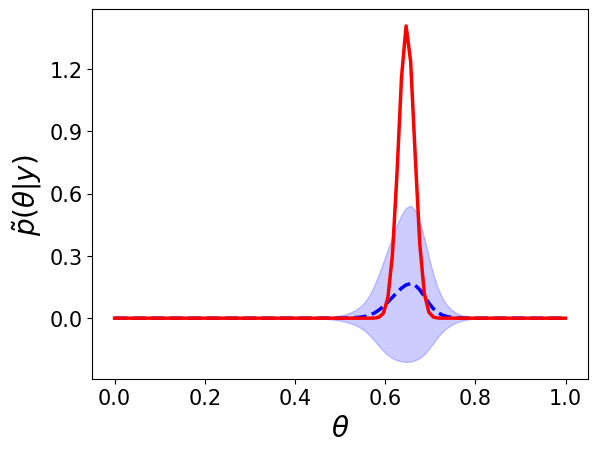}
    \end{subfigure}
    \begin{subfigure}{0.32\textwidth}
        \includegraphics[width=1\textwidth]{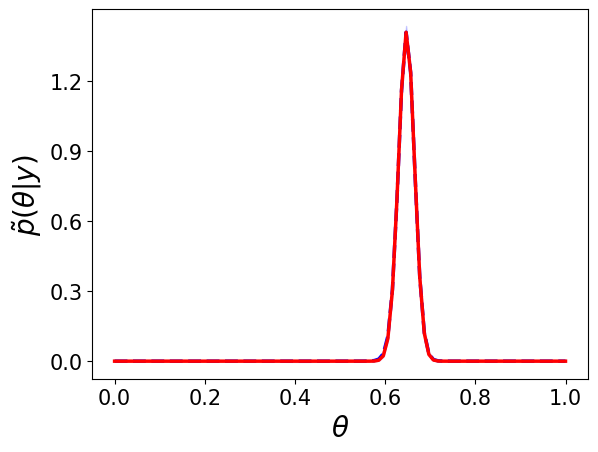}
    \end{subfigure}
    \begin{subfigure}{0.32\textwidth}
        \includegraphics[width=1\textwidth]{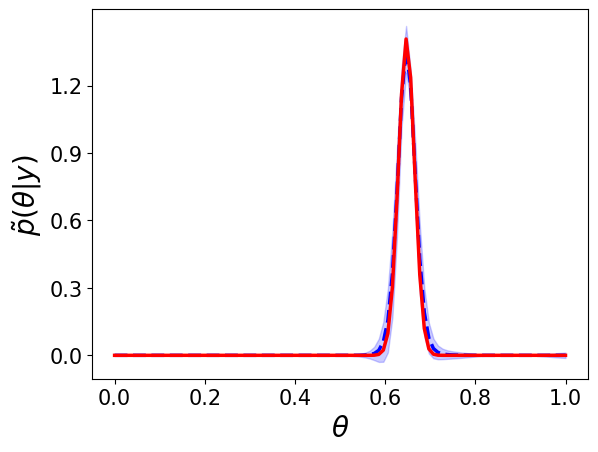}
    \end{subfigure}
    \caption{Illustration with a simulation model $\eta(x, \theta) = \sin(10x - 5\theta)$ where $x \in [0,1]$ and $\theta \in [0, 1]$. Black dots represent the field data replicated twice at the design inputs $(x_1^f, x_2^f, x_3^f, x_4^f, x_5^f) = (0.1, 0.3, 0.5, 0.7, 0.9)$ and generated through $y(x) = \eta\left(x, \theta = \frac{\pi}{5}\right) + \epsilon$, where $\epsilon \sim {\rm N}(0, 0.2^2)$. The top and bottom panels show field and posterior predictions, respectively. Predictions in panels are obtained with a GP emulator built with the simulation data set using LHS (left), $\mathcal{A}_t^p(\cdot)$ (middle), and $\mathcal{A}_t^y(\cdot)$ (right). The blue dashed line shows the prediction mean and the shaded area illustrates one predictive standard deviation from the mean. The red line illustrates the simulation model at $\theta=\frac{\pi}{5}$ (top panels) or the posterior (bottom panels). }
    \label{illustration_post}
\end{figure}

\begin{figure}[t]
    \centering
    \begin{subfigure}{0.32\textwidth}
        \includegraphics[width=1\textwidth]{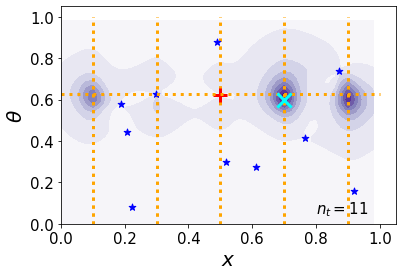}
    \end{subfigure}
    \begin{subfigure}{0.32\textwidth}
        \includegraphics[width=1\textwidth]{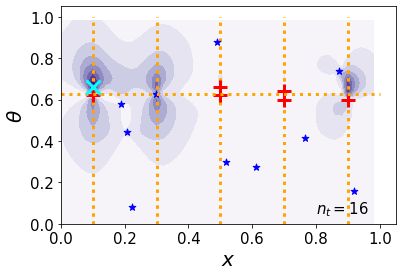}
    \end{subfigure}
    \begin{subfigure}{0.32\textwidth}
        \includegraphics[width=1\textwidth]{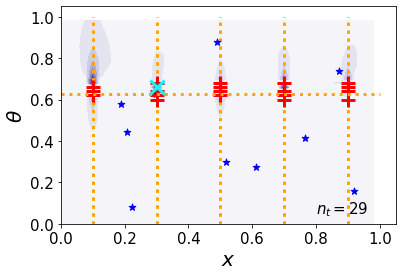}
    \end{subfigure}
    \begin{subfigure}{0.32\textwidth}
        \includegraphics[width=1\textwidth]{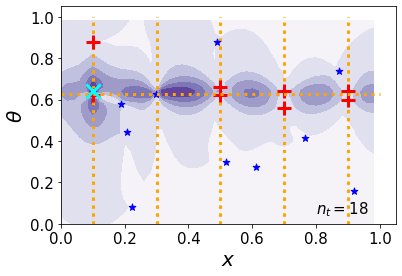}
    \end{subfigure}
    \begin{subfigure}{0.32\textwidth}
        \includegraphics[width=1\textwidth]{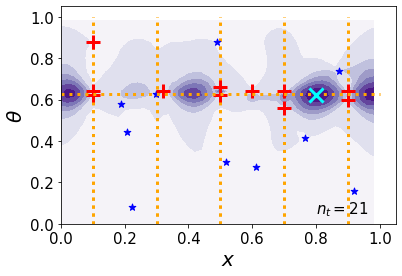}
    \end{subfigure}
    \begin{subfigure}{0.32\textwidth}
        \includegraphics[width=1\textwidth]{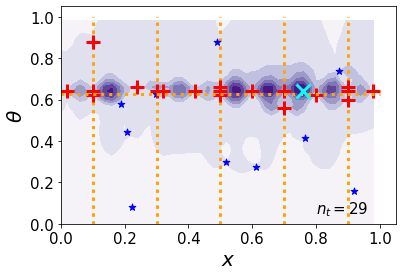}
    \end{subfigure}
    \caption{Acquisition function value surface for three iterations of the sequential procedure via $\mathcal{A}_t^p(\cdot)$ (top panels) and $\mathcal{A}_t^y(\cdot)$ (bottom panels) using Figure~\ref{illustration_post} example. Dark purple indicates lower values and light purple indicates larger values. Markers demonstrate the initial sample (blue star), previously acquired inputs (red plus), and the input minimizing the acquisition function at the current iteration (cyan cross). The orange dotted lines show $\theta = \frac{\pi}{5}$ (horizontal line) and field data design inputs (vertical lines).}
    \label{illustration_eivar}
\end{figure}

As an illustration, consider the simulation model $\eta\left(x, \theta\right)$ presented in  Figure~\ref{illustration_post} with a design input $x$ and a parameter $\theta$. Note that the boldface characters are removed from both $x$ and $\theta$ since they are scalars.
Figure~\ref{illustration_post} shows field predictions (top panels) and posterior predictions (bottom panels) for three different acquisition functions represented by columns. To initialize, $n_0 = 10$ inputs are sampled uniformly from the prior and then $n=20$ simulation outputs are collected with each acquisition function using Algorithm~\ref{alg:oaat}. The emulator returned by the end of the sequential procedure is employed to produce field and posterior predictions as follows. The field predictions are obtained with the emulator mean and variance in \eqref{eq:meanvar_latent} across design inputs paired with $\hat{\theta}$, the estimate of the true value of $\theta$. The parameter estimate $\hat{\theta}$ is iteratively updated to minimize the sum of the squared errors between the field data and its corresponding predictions. The posterior mean and variance are obtained by applying Lemma~\eqref{lemma:UQ} with the associated GP emulator. For example, predictions in the left panels are produced using the simulation data set constructed with LHS and the emulator is not adequate to predict both the field data and posterior truly. In the following, we describe the resulting predictions with the proposed acquisition functions shown in the middle and right panels of Figure~\ref{illustration_post}. Moreover, Figure~\ref{illustration_eivar} illustrates the acquisition function value surface for three iterations, presenting different stages for each function to better explain their behavior. 

The middle panels in Figure~\ref{illustration_post} demonstrate field (top) and posterior (bottom) predictions obtained through the emulator fitted to the simulation data collected with $\mathcal{A}^{p}_t(\cdot)$. The top panels in Figure~\ref{illustration_eivar} illustrate three iterations of the sequential procedure using $\mathcal{A}^{p}_t(\cdot)$. In the calibration space, acquired inputs are concentrated around the (unknown) parameter region of interest, whereas in the design space, $\mathcal{A}^{p}_t(\cdot)$ encourages the selection of design inputs where the field data is collected. The acquisition function $\mathcal{A}^{p}_t(\cdot)$ places the points around the parameter region of interest since the volume of the high-posterior region is small and parameters with zero-posterior density provide almost no information for the behavior of the simulation model near the parameter of interest. Moreover, as demonstrated in Figure~\ref{illustration_post}, focusing only on the regions where the field data is collected improves the posterior prediction dramatically as compared to the one with LHS. Notice that the emulator uncertainty is shrunk towards zero around the field data design inputs since the targeted data collection results in a refined estimate of the field data, with reduced uncertainty in regions where the simulation outputs have been observed. These results are promising especially for high-dimensional design spaces. If one is only interested in accurate and precise inference of calibration parameters through estimating the posterior density, then building independent GP emulators for each design input is much more efficient than building one large GP emulator in high-dimensional spaces (see examples of such emulators in \cite{Gu2016, Huang2020}). The acquisition function $\mathcal{A}^{p}_t(\cdot)$ can still be used with these efficient emulators to construct the simulation data set by restricting the design space to the ones where the physical experiment is conducted. However, since $\mathcal{A}^{p}_t(\cdot)$ does not encourage the exploration of the design space, it does not improve field predictions at unseen design inputs.

The emulator built with the simulation data set collected with $\mathcal{A}^{y}_t(\cdot)$ generates superior field predictions as compared to other approaches (see the right panel in Figure~\ref{illustration_post}). As illustrated in Figure~\ref{illustration_eivar} (bottom panels), early in the sequential procedure, $\mathcal{A}^{y}_t(\cdot)$ encourages the selection of inputs around the design points where the field data is collected to minimize the total uncertainty. Then, it focuses on the entire design space for improved predictions around the parameter of interest. Thus, the emulator uncertainty is reduced not only around the field data design inputs but also over the remaining design space when predicting the field data. Although the posterior predictions obtained with $\mathcal{A}^{p}_t(\cdot)$ match almost perfectly with the ground truth, $\mathcal{A}^{y}_t(\cdot)$ also performs well in terms of predicting the posterior since the simulation data set includes points aligned with field data design inputs as well. In this sense, $\mathcal{A}^{y}_t(\cdot)$ is a way to obtain accurate posterior and field predictions by balancing the exploration of unseen design inputs and exploitation of existing design inputs around the parameter region of interest.

At iteration $t$, acquisition functions $\mathcal{A}^{p}_t(\cdot)$ and $\mathcal{A}^{y}_t(\cdot)$ choose the next input point as the one that minimizes the total uncertainty over the entire parameter and design space, respectively. To evaluate each acquisition function with any candidate input, the expected variance of the posterior is obtained as follows and the derivation is provided in Appendix~\ref{app:3.2}.
\begin{lemma}\label{prop:IEV}
Under the conditions of Lemma~\ref{lemma:UQ}, 
    \begin{align} \label{eq:EIVAR}
        \begin{split}
            & \mathbb{E}_{\eta^* | \mathcal{D}_{t}} \left(\mathbb{V}[p(\y|\thetav) \left| (\mathbf{z}^*, \eta^*) \cup \mathcal{D}_{t} \right] \right) \\
            & \hspace{1in}= \frac{f_\mathcal{N}\left(\y; \, \meanv, \, \frac{1}{2}\Sigmav + \cov\right)}{2^d \pi^{d/2} |\Sigmav|^{1/2}} 
            - \frac{f_\mathcal{N}\left(\y; \, \meanv, \, \frac{1}{2}\left(\Sigmav + \cov + \PHI\right)\right)}{2^d \pi^{d/2} |\Sigmav + \cov - \PHI|^{1/2}},
        \end{split}
    \end{align}
    where the $(i,j)$th element of $\PHI$ is $\frac{\text{\rm cov}_{t}\left(\mathbf{z}_i^f, \mathbf{z}^*\right)\text{\rm cov}_{t}\left(\mathbf{z}_j^f, \mathbf{z}^*\right)}{\varsigma^2_{t}(\mathbf{z}^*) + \upsilon}$ with $\mathbf{z}_i^f = \left({\xb_i^f}^\top, \thetav^\top\right)^\top$ for $i, j = 1, \ldots, d$. 
\end{lemma}
The expected variance of the posterior needs to be integrated over a multi-dimensional parameter and design spaces to obtain Equations~\eqref{EIVAR_post}--\eqref{EIVAR_pred}.
Since the integrals are analytically difficult to compute, they are approximated with a sum over uniformly distributed reference grids $\Theta_{\rm ref}$ and $\mathcal{X}_{\rm ref}$ within the spaces $\Theta$ and $\mathcal{X}$, respectively. In higher-dimensional inputs, the size of a grid covering the input space becomes very large, and one can consider sparse grids, smart sampling techniques, or quadrature schemes for estimating higher dimensional integrals. Alternatively, the reference set could follow a space-filling construction and one can regenerate the space-filling reference set at each iteration to encourage diversity in search.

Notice that the initial term in \eqref{eq:EIVAR} does not depend on $ \mathbf{z}^*$ and we drop this term to efficiently approximate the acquisition criteria. Thus, minimizing $\mathcal{A}_t^p(\cdot)$ in \eqref{EIVAR_post} is efficiently approximated by maximizing  
\begin{align} \label{eq:appEIVAR_post}
    \begin{split}
        &  \frac{1}{|\Theta_{\rm ref}|} \sum_{\thetav \in \Theta_{\rm ref}} p(\thetav)^2 \left(\frac{f_\mathcal{N}\left(\y; \, \meanv, \, \frac{1}{2}\left(\Sigmav + \cov + \PHI\right)\right)}{2^{d}\pi^{d/2}|\Sigmav + \cov - \PHI|^{1/2}}\right).
    \end{split}
\end{align}
The acquisition function $\mathcal{A}_t^y(\cdot)$ considers the expected posterior variance at the estimate $\hat{\thetav}$ integrated over hypothetical design inputs. To do that, the length-($d+1$) vectors $\mathbf{y}^\mathbf{x}$ and $\muv_{t}^\xb(\hat{\thetav})$ and $(d+1) \times (d+1)$ matrices $\Sigmav^\xb$, $\Sv_{t}^\xb(\hat{\thetav})$, and $\phiv_{t}^\xb\left(\hat{\thetav}, \zb^*\right)$ are constructed by augmenting a possible design input $\xb \in \mathcal{X}_{\rm ref}$ onto the existing design inputs $\xb_1^f, \ldots, \xb_d^f$. Then, approximating the minimization of $\mathcal{A}_t^y(\cdot)$ in \eqref{EIVAR_pred} involves maximizing 
\begin{align} \label{eq:appEIVAR_pred}
    \begin{split}
        &  \frac{1}{|\mathcal{X}_{\rm ref}|} \sum_{\xb\in \mathcal{X}_{\rm ref}} p(\hat{\thetav})^2 \left( \frac{f_\mathcal{N}\left(\mathbf{y}^\mathbf{x}; \, \muv_{t}^\xb( \hat{\thetav}), \, \frac{1}{2}\left(\Sigmav^\xb + \Sv_{t}^\xb(\hat{\thetav}) + \phiv_{t}^\xb\left(\hat{\thetav}, \zb^*\right)\right)\right)}{2^{(d+1)}\pi^{(d+1)/2}|\Sigmav^\xb + \Sv_{t}^\xb(\hat{\thetav}) - \phiv_{t}^\xb\left(\hat{\thetav}, \zb^*\right)|^{1/2}}\right).
    \end{split}
\end{align}
Since $y(\xb)$ is unknown at any $\xb \in \mathcal{X}_{\rm ref}$, the emulator mean at $\left(\xb^\top, \hat{\thetav}^\top\right)^\top$ is used as a plug-in estimator. Recall that we replace the difficult numerical optimization with a discrete search by evaluating the acquisition function $\mathcal{A}_t^p(\cdot)$ $\left(\mathcal{A}_t^y(\cdot)\right)$ on a candidate set of inputs $\mathcal{L}_t$. Thus, at each iteration, \eqref{eq:appEIVAR_post} (\eqref{eq:appEIVAR_pred}) is computed for each $\zb^* \in \mathcal{L}_t$, and then the next input point included in the simulation data set is the maximizer from the discrete set of inputs $\mathcal{L}_t$. As an alternative to optimizing over a discrete set of inputs, once the sum-based approximation of the integrals is obtained, one can employ numerical optimization with the off-the-shelf solvers using closed-form derivatives. We note that as inputs are selected with \eqref{eq:appEIVAR_post} (\eqref{eq:appEIVAR_pred}), the covariance matrix $\Sv_{t}(\thetav)$ ($\Sv_{t}^\xb(\hat{\thetav})$) becomes smaller, promoting exploration in regions with higher uncertainty to minimize the aggregated posterior variance.

Our derivations so far assume that the variance $\sigma^2$ of the residual error is known. However, in some cases, the variance term might not be available in advance. Moreover, another component of uncertainty can exist in the form of a model discrepancy. Kennedy and O'Hagan (KOH, \cite{Ohagan2001}) model the field data as the function of a simulation output plus an additional discrepancy term such that $y\left(\xb_i^f\right) = \eta\left(\xb_i^f, \thetav\right) + b\left(\xb_i^f\right) + \epsilon,$
where $b\left(\xb_i^f\right)$ represents the discrepancy or the bias term at the design input $\xb_i^f$. Although the discrepancy term can create an identifiability problem \citep{Bayarri2007, Jenny2014, Plumlee2017, gu2018scaled, Tuo2015}, the KOH framework has been widely used for calibrating simulation models. One can still use the proposed sequential approach when the error variance is unknown and/or in the case of a discrepancy between the simulation output and field data as follows. 

KOH framework assigns a GP emulator as the prior distribution of both the simulation model and discrepancy term. The modular approach \citep{Bayarri2007} determines unknown hyperparameters of an emulator of the simulation model utilizing only the simulation data. Then, the unknown hyperparameters of an emulator of the discrepancy term are obtained by utilizing discrepancy observations generated as the difference of field data and emulator means of a simulation model at the same input values \citep{Bayarri2009}. In parallel to the modular approach, as described in Section~\ref{sec:GP}, an emulator of the simulation model is built using the simulation data set $\mathcal{D}_t$ at each iteration. Then, the unknown discrepancy covariance hyperparameters, denoted by $\thetav^e$, are estimated by maximizing the likelihood. Let $\Sigmav^e$ denote the $d \times d$ covariance matrix of the discrepancy term and the noise term. Hyperparameters $\thetav^e$ determine the structure of the covariance matrix $\Sigmav^e$. For example, when the discrepancy is negligible, $\sigma^2$ is the only hyperparameter that needs to be estimated (e.g., $\theta^e \coloneqq \sigma^2$ and $\Sigmav^e = \sigma^2 \mathbf{I}$). Once the estimates of $\thetav^e$ are obtained at each iteration, Equation~\eqref{eq:appEIVAR_post} can be computed by replacing $\Sigmav$ with the estimate of $\Sigmav^e$. Similarly, in Equation~\eqref{eq:appEIVAR_pred}, the cross-covariance values between any $\xb \in \mathcal{X}_{\rm ref}$ and the existing design inputs are computed using the estimates of $\thetav^e$ to replace $\Sigmav^\xb$ with the estimated covariance matrix.

\section{Experiments}
\label{sec:experiment}
Section~\ref{sec:synthetic} investigates the performance using two synthetic simulation models without and with discrepancy terms. Section~\ref{sec:highdim} examines the performance with higher dimensional input spaces. Section~\ref{sec:covid} demonstrates the proposed acquisition functions with the COVID-19 epidemiological simulation model.

\subsection{Benchmark with Two Synthetic Simulation Models}
\label{sec:synthetic}

We examine the performance of the proposed acquisition functions $\mathcal{A}^p_t(\cdot)$ and $\mathcal{A}^y_t(\cdot)$ abbreviated by $\mathcal{A}^p$ and $\mathcal{A}^y$ using two synthetic simulation models. For comparison, we sample inputs from two space-filling alternatives: uniformly random from a prior and LHS. These approaches are labeled as $\mathcal{A}^{rnd}$ and $\mathcal{A}^{lhs}$ in the experiments. The proposed sequential procedure is implemented under the Python package Parallel Uncertainty Quantification (PUQ) at \hb{{https://github.com/parallelUQ/}} and the code scripts are provided to replicate the examples.

\begin{figure}[ht]
\centering
    \begin{subfigure}{0.4\textwidth}
        \includegraphics[width=1\textwidth]{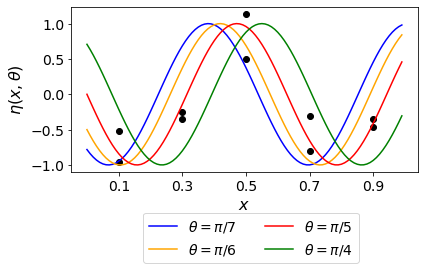}
    \end{subfigure}
    \begin{subfigure}{0.4\textwidth}
        \includegraphics[width=1\textwidth]{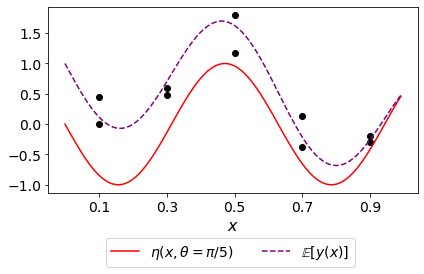}
    \end{subfigure}
    \caption{Illustration of the simulation model with two-dimensional $[\mathcal{X}, \Theta]$ space introduced in Figure~\ref{illustration_post}. In the left panel, lines demonstrate the simulation outputs at $\theta \in \{\frac{\pi}{4}, \frac{\pi}{5}, \frac{\pi}{6}, \frac{\pi}{7}\}$. The red line represents the mean of the field data $\mathbb{E}[y(x)]= \eta\left(x, \theta = \frac{\pi}{5}\right)$ without a discrepancy term. In the right panel, the dashed purple line corresponds to the mean of the field data $\mathbb{E}[y(x)] = \eta\left(x, \theta = \frac{\pi}{5}\right) + b(x)$ with a discrepancy term $b(x) = 1 - \frac{1}{3}x - \frac{2}{3}x^2$. In both panels, black dots represent the field data at equally spaced, five design inputs generated with $y\left(x\right) = \mathbb{E}\left[y\left(x\right)\right] + \epsilon$ using $\epsilon \sim {\rm N}(0, 0.2^2)$.}
    \label{fig:models_sinf}
\end{figure}
\begin{figure}[ht]
\centering
    \begin{subfigure}{0.4\textwidth}
        \includegraphics[width=1\textwidth]{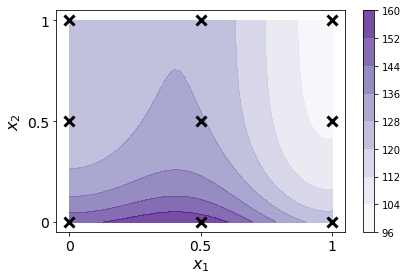}
    \end{subfigure}
    \begin{subfigure}{0.38\textwidth}
        \includegraphics[width=1.05\textwidth]{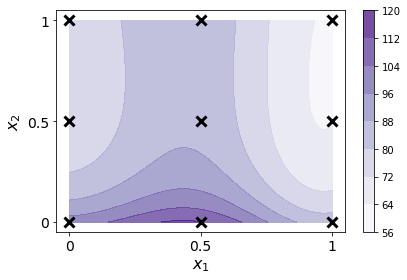}
    \end{subfigure}
    \caption{Illustration of the simulation model with three-dimensional $[\mathcal{X}, \Theta]$ space such that $\eta\left(\xb, \theta\right) = \left(30 + 5 x_1 \sin\left(5 x_1\right)\right)\left(6 \theta + 1 + \exp\left(-5x_2\right)\right)$. The left and right panels illustrate the contour plots of $\mathbb{E}[y(\xb)] = \eta\left(\xb, \theta=0.5\right)$ and $\mathbb{E}[y(\xb)] = \eta\left(\xb, \theta=0.5\right) + b(\xb)$ with $b(\xb) = -50\exp\left(-0.2x_1 - 0.1x_2\right)$, respectively. Cross markers represent nine design inputs where the field data is collected as two replicates of nine points using $y(\xb) = \mathbb{E}[y(\xb)] + \epsilon$ with $\epsilon \sim {\rm N}\left(0, 0.5^2\right)$. }
    \label{fig:models_pritam}
\end{figure}

First, the performance comparison is demonstrated with a sinusoidal simulation model similar to the one used by \cite{Koermer2023}. Throughout the paper, the model has been used for illustration and it includes a one-dimensional design input $x$ and calibration parameter $\theta$ ($q=p=1$) where $x \in \left[0, 1\right]$ and $\theta \in \left[0, 1\right]$. The left panel in Figure~\ref{fig:models_sinf} shows simulation outputs across design space for different parameter values when the discrepancy is negligible. In the right panel, the mean of the field data is demonstrated in the case of discrepancy. For both cases, the data is collected at five unique locations on an equally spaced grid and each observation is replicated twice. The second example comes from \cite{Ranjan2011} with a two-dimensional design input $\xb = \left(x_1, x_2\right)^\top \in [0, 1]^2$ ($q=2$) and a one-dimensional calibration parameter $\theta \in [0,1]$ ($p=1$). Figure~\ref{fig:models_pritam} demonstrates the contour plots of the expected field data without and with discrepancy terms and the locations of nine unique design inputs where the field data is collected. The data generation mechanisms are detailed in the captions of Figures~\ref{fig:models_sinf}--\ref{fig:models_pritam}.

To initialize the sequential procedure, the sample of size $n_0$ is taken from a uniform prior. For the two- and three-dimensional simulation models presented in Figures~\ref{fig:models_sinf}--\ref{fig:models_pritam}, we set $n_0 = 10$ and $n_0 = 30$, respectively, in $[\mathcal{X}, \Theta]$ space. The initial sample size is selected to enable the emulator to sufficiently learn the response surface in the early stages while also permitting enhancements in predictions through subsequent acquisitions. Additional analysis on the initial sample size is provided in Appendix~\ref{app:initial}. The sequential approach is terminated once simulation outputs are collected from $n=90$ and $n=150$ acquired inputs for the two- and three-dimensional functions, respectively. The fixed budget termination criterion is determined based on the point where the best method starts to converge. At each iteration, the candidate list $\mathcal{L}_t$ introduced in lines~5--6 of Algorithm~\ref{alg:oaat} is generated \tcb{to combine exploration using a sample from the prior with exploitation using a sample of parameters paired with field data design inputs}. First, we sample 100 parameters from a uniform prior in $\Theta$ space. Then, \tcb{to account for the influence of the number of field data design inputs}, each unique field data design input is paired with each parameter to construct 500 ($=5 \times 100$) and 900 ($=9 \times 100$) sets of candidate inputs for the first and second models, respectively. These candidate points are generated to facilitate the optimization procedure for exploiting existing field data design inputs. In addition, another 500 and 900 inputs are randomly sampled from the prior in $[\mathcal{X}, \Theta]$ space to allow exploration for the first and second functions, respectively. \tcb{We opt for a 50\%-50\% split to ensure equal representation for exploration and exploitation.} Therefore, at iteration $t$, the size of the candidate list is $|\mathcal{L}_t| = 1000$ and $|\mathcal{L}_t| = 1800$ for the first and second functions. \tcb{The size and construction of the candidate list depend on various experimental characteristics, such as the number of field data design inputs and the dimension of the input space. As dimensionality increases, practitioners can adjust the candidate list based on their computational budget and experimental setup.}

The performance comparison is summarized over 30 replications. At each replication, the initial design of size $n_0$ is randomly chosen from a uniform prior. The same initial sample is used for all methods $\mathcal{A}^p$, $\mathcal{A}^y$, $\mathcal{A}^{rnd}$, and $\mathcal{A}^{lhs}$ for a fair comparison. Similarly, the field data is rerandomized at each replication and the same field data is used across different methods within each replication. As a performance metric, we compute the mean absolute difference ${\rm MAD}^p$ between the estimated posterior and the true posterior $\tilde{p}(\thetav|\y)$ at unseen calibration parameters and the mean absolute difference ${\rm MAD}^y$ between the estimated field data and the true field data at unseen design inputs. To do that, we generate a set of reference calibration parameters $\Theta_{\rm ref}$ and a set of reference design inputs $\mathcal{X}_{\rm ref}$ and compute the performance metrics via ${\rm MAD}_t^p = \frac{1}{|\Theta_{\rm ref}|}\sum_{\thetav \in \Theta_{\rm ref}} |\tilde{p}(\thetav|\y) - \hat{p}_t(\thetav|\y)|$ and ${\rm MAD}_t^y = \frac{1}{|\mathcal{X}_{\rm ref}|}\sum_{\xb \in \mathcal{X}_{\rm ref}} |y(\xb) - \hat{y}_t(\xb)|$ at each iteration $t$. Here, $\hat{p}_t(\thetav|\y)$ is the posterior prediction obtained with Equation~\eqref{expectedpostfinal} and $\hat{y}_t(\xb)$ is the field prediction obtained using the emulator mean in Equation~\eqref{eq:meanvar_latent} at a given design input $\xb$ paired with $\hat{\thetav}$. For the two-dimensional function, both $\Theta_{\rm ref}$ and $\mathcal{X}_{\rm ref}$ are generated from 100 equally spaced points in $[0, 1]$. For the three-dimensional example, $\Theta_{\rm ref}$ is generated from 100 equally spaced points in $[0, 1]$ and $\mathcal{X}_{\rm ref}$ is generated in a two-dimensional grid of $20^2$ points. In addition, the reference sets $\Theta_{\rm ref}$ and $\mathcal{X}_{\rm ref}$ are used to approximate the acquisition function values in Equations~\eqref{eq:appEIVAR_post}--\eqref{eq:appEIVAR_pred}. For all the experiments presented in this section, the calibration parameter estimate, $\hat{\thetav}$, is updated at each iteration by minimizing the sum of the squared errors between the field data and field prediction\tcb{, as it is one of the most common parameter estimation techniques}. \tcb{Alternative techniques, such as maximum likelihood estimation and $\chi^2$ minimization, can also be employed. Additionally, for real-world examples with highly nonlinear response surfaces, field scientists may be aware of specific optimization techniques better suited to parameter estimation in their applications.} For the experiments with a discrepancy term, in addition to $\hat{\thetav}$, GP hyperparameters $\thetav^e$ for the discrepancy term are also estimated. We use a covariance form $\Sigmav^e$ defined by $\Sigmav^e_{i,j} = \sigma_\varepsilon^2 \delta_{i=j} + \sigma_b^2 \exp\left(-\lambda\left|\left|\xb_i^f - \xb_j^f\right|\right|_1\right)$ for $i, j = 1, \ldots, d$ and the maximum likelihood estimates of hyperparameters $\thetav^e = (\sigma_\varepsilon^2, \sigma_b^2, \lambda)^\top$ are obtained at each iteration. Moreover, for the examples with a discrepancy term, we investigate the quality of acquired parameters via the interval score, instead of comparing the methods with the ${\rm MAD}^p$ metric due to unknown $\Sigmav^e$. To do this, we compute the interval score $S_\alpha(l, u; a)$ \citep{gneiting2007strictly} for each method and replicate via
\begin{gather}\label{eq:interval_score}
   S_\alpha(l, u; a) = (u-l) + \frac{2}{\alpha} (l - a) \mathbbm{1}\{{a < l}\} + \frac{2}{\alpha}(a - u)  \mathbbm{1}\{{a > u}\}
\end{gather}
where $l$ and $u$ represent the quantiles of acquired parameters at level $\frac{\alpha}{2}$ and $(1-\frac{\alpha}{2})$, respectively, and $\mathbbm{1}$ is an indicator function. To assess whether the best-fit parameter falls within the range of acquired parameters, we substitute the least-squares fit parameter value in place of $a$ and use $\alpha = 0.10$ in the experiments. The interval score helps address the width of acquired parameters and the coverage of the best-fit parameterization.

\begin{figure}[ht]
\centering
    \begin{subfigure}{0.32\textwidth}
        \includegraphics[width=1\textwidth]{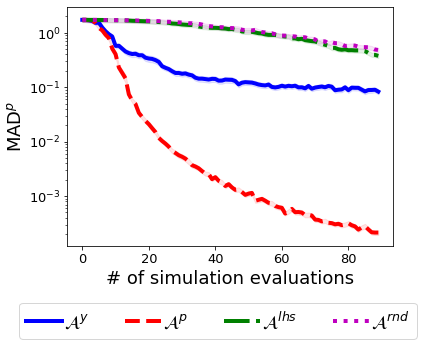}
    \end{subfigure}
    \begin{subfigure}{0.32\textwidth}
        \includegraphics[width=1\textwidth]{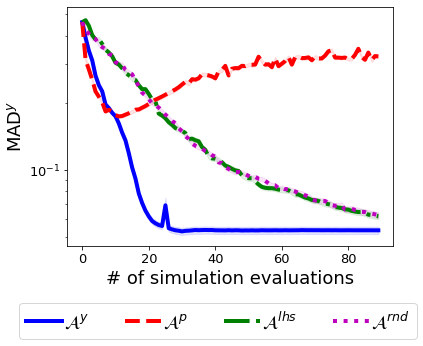}
    \end{subfigure}
    \begin{subfigure}{0.32\textwidth}
        \includegraphics[width=1\textwidth]{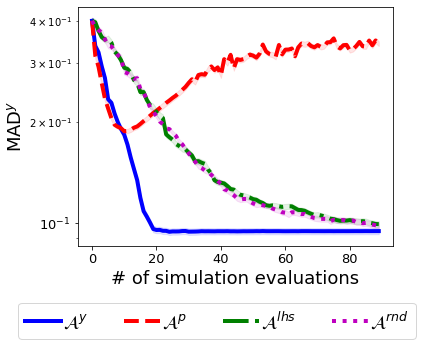}
    \end{subfigure}
    \caption{Comparison of different acquisition functions using the two-dimensional simulation model in Figure~\ref{fig:models_sinf}. The left panel compares the accuracy of posterior predictions (without a discrepancy term). The middle (without a discrepancy term) and the right panels (with a discrepancy term) compare the accuracy of field predictions.}
    \label{sinf_pred_res}
\end{figure}
\begin{figure}[ht]
\centering
    \begin{subfigure}{0.32\textwidth}
        \includegraphics[width=1\textwidth]{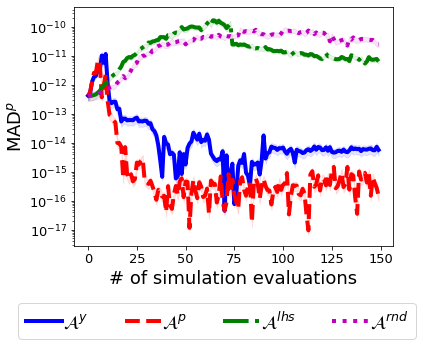}
    \end{subfigure}
    \begin{subfigure}{0.32\textwidth}
        \includegraphics[width=1\textwidth]{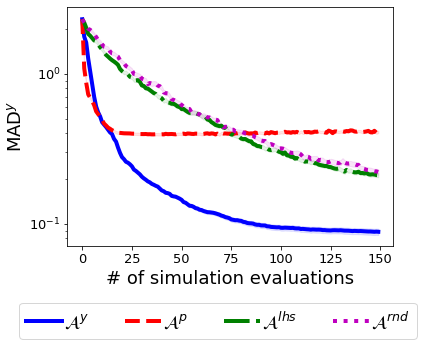}
    \end{subfigure}
    \begin{subfigure}{0.32\textwidth}
        \includegraphics[width=1\textwidth]{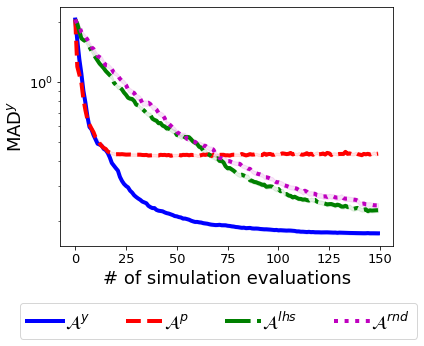}
    \end{subfigure}
    \caption{Comparison of different acquisition functions using the three-dimensional simulation model in Figure~\ref{fig:models_pritam}. The left panel compares the accuracy of posterior predictions (without a discrepancy term). The middle (without a discrepancy term) and the right panels (with a discrepancy term) compare the accuracy of field predictions.}
    \label{pritam_pred_res}
\end{figure}
\begin{figure}[ht]
\centering
    \begin{subfigure}{0.4\textwidth}
        \includegraphics[width=1\textwidth]{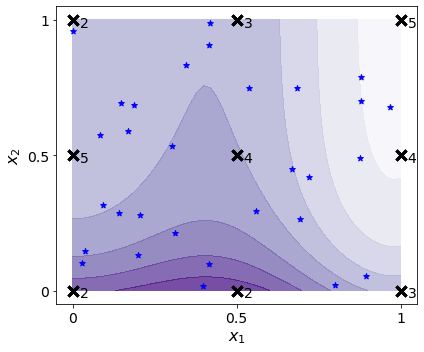}
    \end{subfigure}
    \begin{subfigure}{0.4\textwidth}
        \includegraphics[width=1\textwidth]{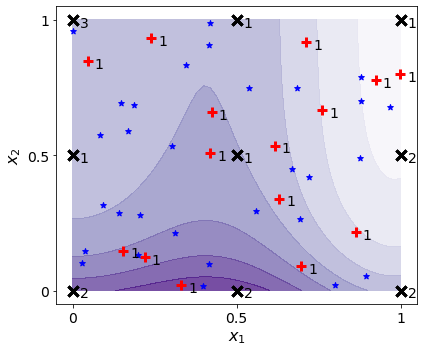}
    \end{subfigure}
    \caption{Illustration of design inputs for 30 acquired inputs collected with $\mathcal{A}^p$ (left panel) and $\mathcal{A}^y$ (right panel) using the three-dimensional simulation model. Blue stars illustrate 30 samples used to initiate the proposed procedure. The cross markers (field data design inputs) and red plus markers (unseen design inputs) demonstrate acquired design inputs. The numbers next to the markers show the number of times each design input is included in the simulation data set.}
    \label{pritam_loc}
\end{figure}
Figures~\ref{sinf_pred_res}--\ref{pritam_pred_res} summarize the performance metrics ${\rm MAD}^p$ and ${\rm MAD}^y$ for the two- and three-dimensional simulation models, respectively. The acquisition function $\mathcal{A}^p$ is superior to its competitors in terms of predicting the posterior as shown in the left panels. However, especially for the first function, $\mathcal{A}^p$ behaves poorly in comparison to other acquisition functions in terms of field predictions. The acquisition function $\mathcal{A}^p$ focuses on the exploitation of regions around existing design inputs and it does not encourage the exploration of the design space $\mathcal{X}$ since unseen design points do not provide any additional value for posterior estimation. As a result, the simulation data set is dominated by the field data design inputs (especially during later stages of the sequential procedure), and the emulator's hyperparameters are tuned based on this data, increasing field prediction error (see middle and right panels in Figure~\ref{sinf_pred_res}). On the other hand, the acquisition function $\mathcal{A}^y$ encourages the selection from the entire space $\mathcal{X}$ around the calibration parameter of interest for improved field predictions. Figure~\ref{pritam_loc} demonstrates the design inputs included in the simulation data set constructed by $\mathcal{A}^p$ and $\mathcal{A}^y$ for a single replicate of the second simulation model. $\mathcal{A}^p$ explores the calibration space located on the field data design inputs to minimize the aggregated uncertainty of the posterior over the parameter space. As a result, $\mathcal{A}^p$ distributes the parameters around both high and low posterior regions to better learn the posterior and covers the parameter region of interest well enough by not wasting any computational resources for simulation evaluations outside of the region of interest. On the other hand, $\mathcal{A}^y$ encourages filling the design space to accurately predict the field data while exploiting the field data design inputs to reduce the variance of the posterior at the parameter estimate. Therefore, $\mathcal{A}^y$ achieves not only the best field prediction accuracy in both examples but also more accurate posterior predictions than the space-filling alternatives. 

\begin{table}[h]
    \centering
    \begin{tabular}{c|c|cccc|cccc}
        & & \multicolumn{4}{c}{Example in Figure~\ref{fig:models_sinf}} & \multicolumn{4}{c}{Example in Figure~\ref{fig:models_pritam}} \\ \hline
        & \diagbox{Metric}{Method} & $\mathcal{A}^{y}$ & $\mathcal{A}^{p}$ & $\mathcal{A}^{lhs}$ & $\mathcal{A}^{rnd}$ & $\mathcal{A}^{y}$ & $\mathcal{A}^{p}$ & $\mathcal{A}^{lhs}$ & $\mathcal{A}^{rnd}$ \\ \hline
     \multirow{2}{*}{without discrepancy} & ${\rm MAD}^p$  & 14 & 11 & 81 & 90 & 38 & 19 & NA & NA \\
                           & ${\rm MAD}^y$  & 21 & NA & 89 & 90 & 29 & NA & 138 & 150 \\
      with discrepancy & ${\rm MAD}^y$  & 20 & NA & 90 & 88 & 37 & NA & 129 & 150 \\ \hline
    \end{tabular}
    \caption{Number of acquired inputs required to achieve a certain accuracy level. ``NA'' means the associated method is not able to attain the desired accuracy level.}
    \label{tab:data_sample}
\end{table}
In Figure~\ref{pritam_pred_res}, the posterior prediction error increases with $\mathcal{A}^{lhs}$ and $\mathcal{A}^{rnd}$ during the early stages of the sequential process since the posterior is constantly predicted with a large positive bias. To see the number of samples required for $\mathcal{A}^{lhs}$ and $\mathcal{A}^{rnd}$ to reach the same posterior prediction accuracy level with $\mathcal{A}^{p}$, we increased the sample size and found that about 500 additional inputs are needed with $\mathcal{A}^{lhs}$ and $\mathcal{A}^{rnd}$. In addition, Table~\ref{tab:data_sample} presents the number of simulation evaluations required to achieve a specific accuracy level for two- and three-dimensional examples. For the example presented in Figure~\ref{fig:models_sinf}, for instance, to attain the particular ${\rm MAD}^p$ level achieved by $\mathcal{A}^{rnd}$ with 90 acquired inputs, $\mathcal{A}^{y}$, $\mathcal{A}^{p}$, and $\mathcal{A}^{lhs}$ require 14, 11, and 81 acquisitions, respectively. Therefore, the targeted sampling approach with $\mathcal{A}^{p}$ and $\mathcal{A}^{y}$ can be useful especially when working with expensive simulation models since $\mathcal{A}^{p}$ and $\mathcal{A}^{y}$ require fewer number of simulation evaluations than the space-filling approaches to achieve the same level of calibration objective. 
\begin{table}[h]
    \centering
    \begin{tabular}{c|cccc|cccc}
         & \multicolumn{4}{c}{Example in Figure~\ref{fig:models_sinf}} & \multicolumn{4}{c}{Example in Figure~\ref{fig:models_pritam}} \\ \hline
         \diagbox{Metric}{Method} & $\mathcal{A}^{y}$ & $\mathcal{A}^{p}$ & $\mathcal{A}^{lhs}$ & $\mathcal{A}^{rnd}$ & $\mathcal{A}^{y}$ & $\mathcal{A}^{p}$ & $\mathcal{A}^{lhs}$ & $\mathcal{A}^{rnd}$ \\ \hline
         Average Interval Score  & 0.11 & 0.08 & 0.89 & 0.89 & 0.07 & 0.05 & 0.89 & 0.90 \\ \hline
    \end{tabular}
    \caption{Comparison of different acquisition functions using the examples in Figures~\ref{fig:models_sinf}--\ref{fig:models_pritam} in the case of discrepancy. The average interval score is computed across 30 replicates for each acquisition function.}
    \label{fig:interval_score}
\end{table}
The goal of $\mathcal{A}^p$ is to better infer the calibration parameters through learning the posterior density. To illustrate this in the presence of discrepancy as well, Table~\ref{fig:interval_score} examines the quality of acquired parameters by averaging the interval score across 30 replicates of each method. In both examples, the small interval score indicates that $\mathcal{A}^p$ is able to target the high posterior region and collect parameters concentrated around the best-fit parameterization. \tcb{Therefore, if the goal is to obtain an accurate and precise estimate of the posterior rather than field predictions, we recommend using the acquisition function $\mathcal{A}^p$. However, in some situations, $\mathcal{A}^y$ may be more advantageous if the objective is to exploit the region around the best-fit parameterization rather than to learn the entire posterior, especially when the high posterior region is large.} Moreover, deciding whether the discrepancy is negligible or not when employing $\mathcal{A}^p$ or $\mathcal{A}^y$ is an important question similar to the other Bayesian calibration procedures. In such a case, a practitioner can consult with the domain scientist's opinion to determine whether known discrepancies exist between the model and the field data. Additionally, one can perform sensitivity analysis techniques to identify whether including potential discrepancies improves the model's performance; see \cite{Sung2024} for a recent review on calibration and an extensive discussion on the discrepancy.

\subsection{Benchmark with High Dimensional Inputs}
\label{sec:highdim}

\begin{figure}[ht]
\centering
    \begin{subfigure}{1\textwidth}
        \includegraphics[width=1\textwidth]{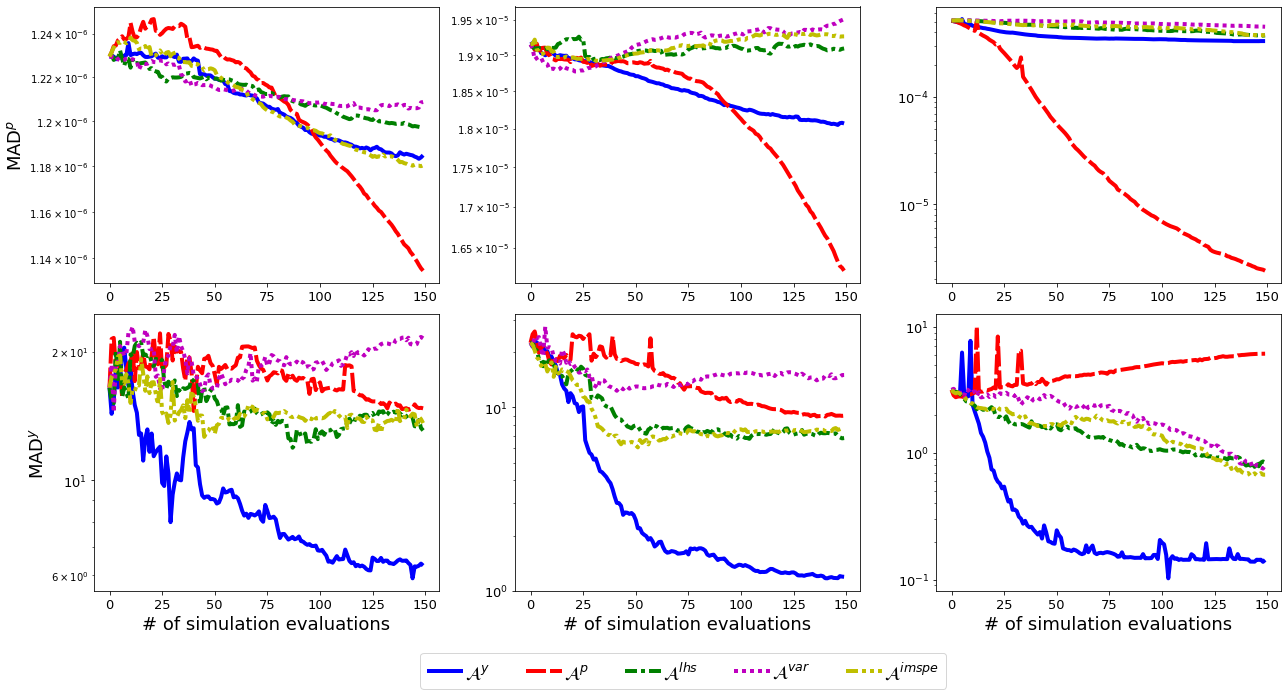}
    \end{subfigure}
    \caption{Comparison of different acquisition functions using the higher-dimensional input settings with $q=2$, $p=10$ (left), $q=6$, $p=6$ (middle), $q=10$, $p=2$ (right). The top panel compares the accuracy of posterior predictions and the bottom panel compares the accuracy of field predictions.}
    \label{fig:highdim_summary}
\end{figure}

This section investigates the performance of the proposed acquisition functions using higher-dimensional input spaces. To understand the effectiveness of the proposed approaches across different configurations of $q$-dimensional design input $\xb$ and $p$-dimensional calibration parameter $\thetav$, we maintain the input dimension at $q + p = 12$ and consider three different scenarios of $q$ and $p$: $q=2$, $p=10$; $q=6$, $p=6$; and $q=10$, $p=2$. Details of the data generation mechanism are given in Appendix~\ref{app:highdim}. In addition to the methods outlined in Section~\ref{sec:synthetic}, we include two common acquisition functions, namely $\mathcal{A}^{var}$ and $\mathcal{A}^{imspe}$, to highlight their differences from the proposed approaches tailored for calibration. While $\mathcal{A}^{var}$ selects the input with the highest emulation variance, $\mathcal{A}^{imspe}$ acquires an input to minimize the aggregated emulation variance. Both functions are typically employed to build globally accurate emulators of simulation models. For the sake of completeness, the implementation details of $\mathcal{A}^{var}$ and $\mathcal{A}^{imspe}$ are provided in Appendix~\ref{app:highdim}. We note that $\mathcal{A}^{rnd}$ is excluded from the results since it performs comparably to $\mathcal{A}^{lhs}$. For all the examples, the sequential procedure terminates once $n=150$ inputs and the associated simulation outputs are collected. Figure~\ref{fig:highdim_summary} summarizes the performance metrics ${\rm MAD}^p$ and ${\rm MAD}^y$ across 10 replications, similar to the experiments in Section~\ref{sec:synthetic}. To gain insights into the differences between the inputs acquired by each method, we compute the width of the interval between the $5 \%$ and $95 \%$ quantiles of each design input $x_i$, $i = 1, \ldots, q$, at each replicate. We also measure the interval score for each calibration parameter $\theta_i$, $i = 1, \ldots, p$, via \eqref{eq:interval_score} to see both the coverage of the best-fit parameterization and the width of the interval. Table~\ref{tab:interval_score_high} provides the width of design inputs and the interval score of parameters averaged across 10 replicates.
\begin{table}[ht]
\footnotesize
    \centering
    \setlength{\tabcolsep}{3pt}
    \begin{tabular}{cccccc|cccccc|cccccc}
          & $\mathcal{A}^y$ & $\mathcal{A}^p$ & $\mathcal{A}^{lhs}$ & $\mathcal{A}^{var}$ & $\mathcal{A}^{imspe}$ & 
          & $\mathcal{A}^y$ & $\mathcal{A}^p$ & $\mathcal{A}^{lhs}$ & $\mathcal{A}^{var}$ & $\mathcal{A}^{imspe}$ & 
          & $\mathcal{A}^y$ & $\mathcal{A}^p$ & $\mathcal{A}^{lhs}$ & $\mathcal{A}^{var}$ & $\mathcal{A}^{imspe}$\\ \hline
       $x_1$          & 0.87 & 0.09 & 0.89 & 0.94 & 0.81 & $x_1$       & 0.85 & 0.00 & 0.89 & 0.95 & 0.85 & $x_1$      & 0.78 & 0.00 & 0.89 & 0.97 & 0.87 \\
       $x_2$          & 0.90 & 0.09 & 0.89 & 0.95 & 0.82 & $x_2$       & 0.86 & 0.00 & 0.89 & 0.95 & 0.85 & $x_2$      & 0.79 & 0.00 & 0.89 & 0.96 & 0.87 \\
       $\theta_1$     & 0.69 & 0.78 & 0.89 & 0.98 & 0.88 & $x_3$       & 0.86 & 0.00 & 0.89 & 0.95 & 0.84 & $x_3$      & 0.78 & 0.00 & 0.89 & 0.96 & 0.87 \\
       $\theta_2$     & 0.63 & 0.79 & 0.89 & 0.98 & 0.88 & $x_4$       & 0.83 & 0.00 & 0.89 & 0.95 & 0.86 & $x_4$      & 0.78 & 0.00 & 0.89 & 0.96 & 0.86 \\ 
       $\theta_3$     & 0.69 & 0.80 & 0.89 & 0.97 & 0.87 & $x_5$       & 0.86 & 0.00 & 0.89 & 0.95 & 0.86 & $x_5$      & 0.79 & 0.00 & 0.89 & 0.96 & 0.86 \\
       $\theta_4$     & 0.64 & 0.79 & 0.89 & 0.98 & 0.87 & $x_6$       & 0.85 & 0.00 & 0.89 & 0.95 & 0.84 & $x_6$      & 0.79 & 0.00 & 0.89 & 0.96 & 0.86 \\
       $\theta_5$     & 0.70 & 0.79 & 0.89 & 0.98 & 0.85 & $\theta_1$  & 0.45 & 0.76 & 0.89 & 0.98 & 0.91 & $x_7$      & 0.78 & 0.00 & 0.89 & 0.96 & 0.87 \\
       $\theta_6$     & 0.65 & 0.80 & 0.89 & 0.97 & 0.87 & $\theta_2$  & 0.45 & 0.80 & 0.89 & 0.98 & 0.91 & $x_8$      & 0.80 & 0.00 & 0.89 & 0.96 & 0.87 \\
       $\theta_7$     & 0.61 & 0.78 & 0.89 & 0.98 & 0.86 & $\theta_3$  & 0.48 & 0.77 & 0.89 & 0.98 & 0.91 & $x_9$      & 0.79 & 0.00 & 0.89 & 0.95 & 0.88 \\
       $\theta_8$     & 0.67 & 0.77 & 0.89 & 0.98 & 0.88 & $\theta_4$  & 0.45 & 0.79 & 0.89 & 0.98 & 0.90 & $x_{10}$   & 0.80 & 0.00 & 0.89 & 0.96 & 0.86 \\
       $\theta_9$     & 0.65 & 0.79 & 0.89 & 0.98 & 0.87 & $\theta_5$  & 0.46 & 0.72 & 0.89 & 0.98 & 0.90 & $\theta_1$ & 0.14 & 0.86 & 0.89 & 0.99 & 0.93 \\
       $\theta_{10}$  & 0.64 & 0.77 & 0.89 & 0.98 & 0.87 & $\theta_6$  & 0.51 & 0.77 & 0.89 & 0.98 & 0.90 & $\theta_2$ & 0.14 & 0.88 & 0.89 & 0.99 & 0.93 \\ \hline
    \end{tabular}
    \caption{The width of design inputs and the interval score of parameters selected with different acquisition functions using the examples with 12-dimensional inputs: $q=2$, $p=10$ (left), $q=6$, $p=6$ (middle), $q=10$, $p=2$ (right).}
    \label{tab:interval_score_high}
\end{table}

While $\mathcal{A}^y$ results in the lowest interval score across all calibration parameters, $\mathcal{A}^p$ consistently attains the narrowest width across all design inputs. Since $\mathcal{A}^y$ tightly acquires around the plausible parameter region and explores the design space well, it achieves the most precise predictions of field data. We find that the width of design inputs acquired with $\mathcal{A}^{p}$ is zero in many cases, indicating that $\mathcal{A}^{p}$ acquires design points from which the field data is collected. Moreover, compared to $\mathcal{A}^{lhs}$, $\mathcal{A}^{var}$, and $\mathcal{A}^{imspe}$, $\mathcal{A}^p$ better constraints the parameter region of interest. As a result, $\mathcal{A}^p$ has a far superior posterior predictive performance. $\mathcal{A}^{var}$ tends to select points at the boundary where the emulation uncertainty is higher; thus, it has the highest interval score and does not perform well in all cases. On the other hand, $\mathcal{A}^{imspe}$ performs relatively better since it fills in interior regions. $\mathcal{A}^{imspe}$ has an overall performance that is comparable to $\mathcal{A}^{lhs}$ for calibration since both $\mathcal{A}^{imspe}$ and $\mathcal{A}^{lhs}$ do not perform a targeted sampling. $\mathcal{A}^{imspe}$ and $\mathcal{A}^{lhs}$ place the inputs far from the region of interest and these inputs provide little information about the behavior of the simulation model near the calibration region of interest. We note that although increasing the dimension of the design and parameter space does not affect the proposed methodology, it results in additional computational costs due to higher-dimensional integrals. One can use alternative sampling methods mentioned in Section~\ref{sec:acquisition} to address the curse of dimensionality for approximating higher-dimensional integrals. Additionally, since the proposed approaches require fewer simulation evaluations than space-filling approaches to achieve the same level of calibration goal, the computational expense is less of a concern, especially for expensive simulation models.

\subsection{Application to an Epidemiological Simulation Model}
\label{sec:covid}

We illustrate the proposed design strategy on a real data example of the COVID-19 epidemiological simulation model. The simulation outputs are generated by the COVID-19 differential equation-based simulation model presented in \cite{Yang2020}. Yang et al. demonstrate how simulation outputs (e.g., forecasted daily admissions and census hospitalizations, daily and census ICU hospitalizations, confirmed cases, and deaths) helped decision makers to decide on whether the community mitigation measures should be enhanced or relaxed and to guide public policies throughout the COVID-19 epidemic in a large US city Austin, Texas. In their simulation-based optimization model, Yang et al. discard the simulation outputs that are inconsistent with observed data using the coefficient of determination (i.e., $r^2$) as a metric to evaluate the quality of simulation outputs, and then the optimization model is built with the filtered simulation data. In parallel to this, \cite{Surer2021} propose a filtering approach to remove unrealistic simulation outputs from the simulation data and show that the calibration with the filtered simulation data reduces the uncertainty in the parameters and the resulting predictions of the COVID-19 model. One downside of such a filtering approach in both procedures is that deciding on plausible simulation outputs based on a priori threshold value may result in oversampling or undersampling. In this sense, our approach can be considered as a systematic way of selecting simulation outputs that are consistent with observations. Therefore, the output of the proposed sequential approach (either the simulation data or the emulator) can serve as a substitute for the filtering procedures in various settings.
\begin{figure}[ht]
\centering
    \begin{subfigure}{0.45\textwidth}
        \includegraphics[width=1\textwidth]{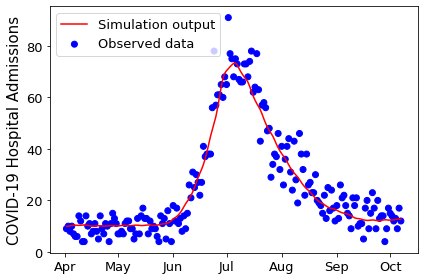}
    \end{subfigure}
    \begin{subfigure}{0.45\textwidth}
        \includegraphics[width=1\textwidth]{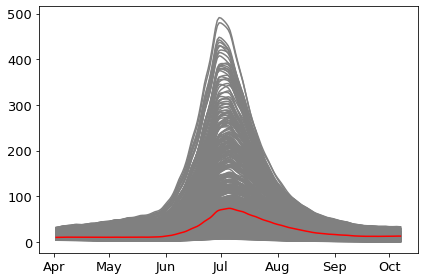}
    \end{subfigure}
    \caption{Illustration of the COVID-19 data from April 1 through October 6, 2020 in Austin. The left panel shows the daily COVID-19 hospital admissions. The red line corresponds to the simulation output at the best-fit parameter $\breve{\thetav}$ during the calibration period. In the right panel, gray lines are simulation outputs across design points paired with 500 different settings of the calibration parameter sampled using LHS.}
    \label{fig:covid_model}
\end{figure}
    
\begin{figure}[ht]
    \centering
    \includegraphics[width=\textwidth]{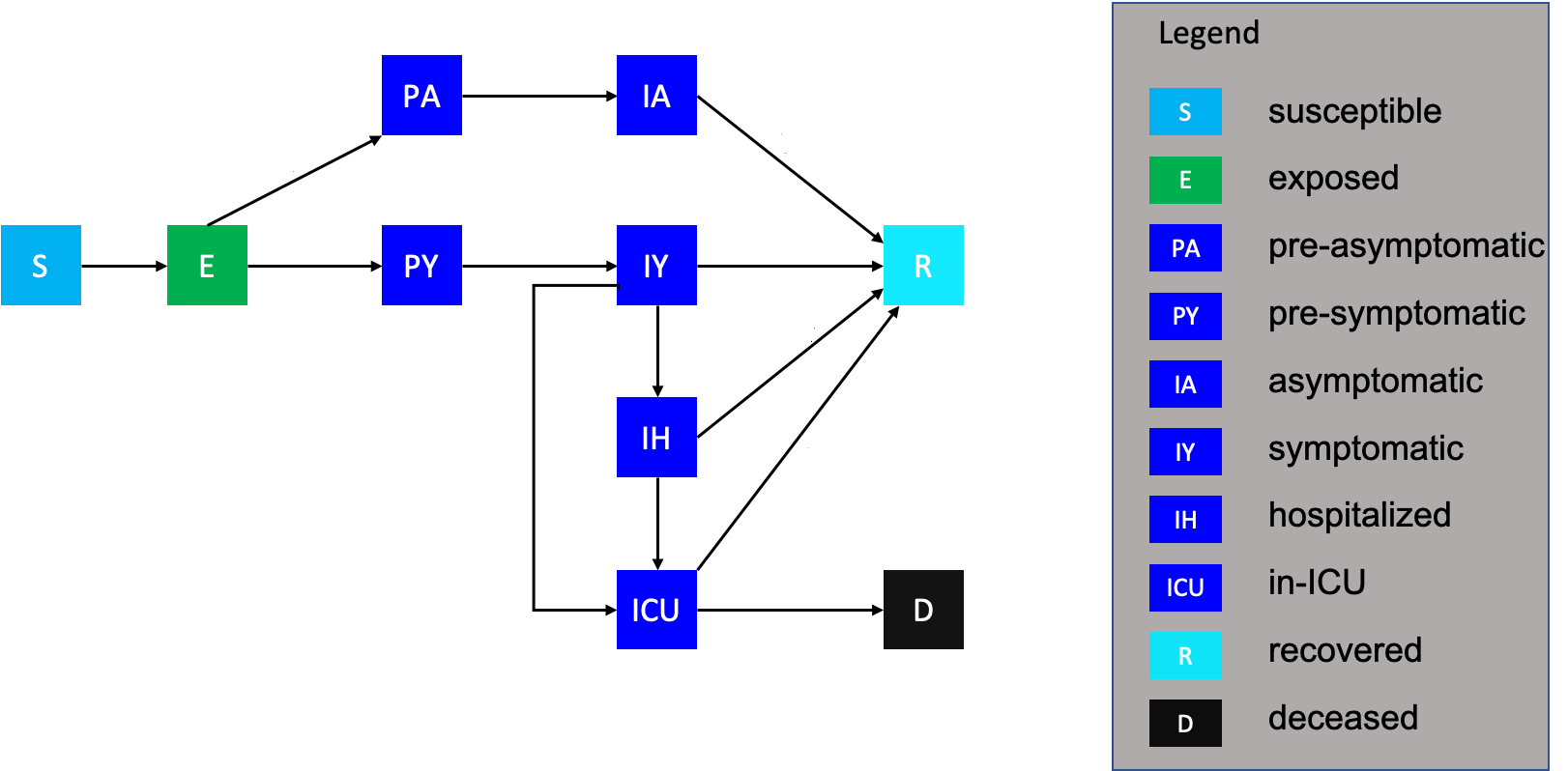}
    \caption{Diagram of the enhanced SEIR-style model comprising ten compartments \citep{Yang2020}.}
    \label{fig:SEIR_Diagram}
\end{figure}

Figure~\ref{fig:covid_model} shows the COVID-19 hospital admissions from April 1 through October 6, 2020 in Austin. Since this period is used to calibrate the simulation model in \cite{Yang2020} to initiate their projection period afterward, we consider the simulation outputs during the same period in our case study. The epidemiological simulation model is an enhanced Susceptible-Exposed-Infectious-Removed (SEIR)--style model comprising ten compartments provided in Figure~\ref{fig:SEIR_Diagram} and the IH (hospitalized) compartment represents the quantity of interest in this case study. We consider the uncertain parameter $\thetav = (1/\sigma_I, \omega_A, 1/\gamma_Y, 1/\gamma_A)^\top$ that affects epidemiological transition dynamics between and within compartments, and the definitions are provided in Table~\ref{tab:covid_params}. Rather than using the rates $\sigma_I$, $\gamma_Y$, and $\gamma_A$, we consider the inverse $1/\sigma_I$, $1/\gamma_Y$, and $1/\gamma_A$, which correspond to duration in days, since the prior information is provided by the experts for the distributions of durations. In \cite{Yang2020}, the best-fit parameterization for the epidemiological parameter $\thetav$ is obtained at $\breve{\thetav} = (2.9, 0.66, 4, 4)^\top$ via the least-squares estimation. In our setting, each day from April 1 through October 6, 2020 (a total of 189 days) corresponds to a design input $x$. The red line in Figure~\ref{fig:covid_model} shows the simulation output across design inputs paired with $\breve{\thetav}$. Since it shows an agreement with the observed hospitalizations, we use this model output as the ``true'' model in our case study. In addition, we consider $\breve{\thetav}$ as a plug-in estimator when selecting inputs with $\mathcal{A}^y$ to evaluate the performance when an accurate estimate of the parameter of interest is available to the practitioner (i.e., $\hat{\thetav} = \breve{\thetav}$ in \eqref{eq:appEIVAR_pred} throughout the procedure). Moreover, we allow $\hat{\thetav}$ to be updated after each active learning acquisition similar to the experiments presented in Sections~\ref{sec:synthetic}--\ref{sec:highdim}, denoting the corresponding result as $\mathcal{\hat{A}}^y$. \tcb{In addition to $\mathcal{A}^y$, $\mathcal{\hat{A}}^y$, and $\mathcal{A}^p$, we include $\mathcal{A}^{lhs}$, $\mathcal{A}^{var}$, and $\mathcal{A}^{imspe}$ in the benchmark, as in the experiments in Section~\ref{sec:highdim}, and exclude $\mathcal{A}^{rnd}$ since its performance is similar to $\mathcal{A}^{lhs}$.} The field data is observed at 13 equally spaced days starting from April 1, 2020, with 15-day intervals (i.e., $d=13$) since a 15-day interval is enough to capture the changes in hospital admissions. Each design input and parameter is scaled to $[0, 1]$ to simplify the integrals. For each of the design inputs, the observed data is simulated as $y\left(x\right) = \eta\left(x, \thetav = \breve{\thetav}\right) + \epsilon$ where $\epsilon \sim {\rm N}(0, 5^2)$. We perform 30 replications with $n_0 = 50$ and $n = 150$. The set $\Theta_{\rm ref}$ is constructed with $500$ points using LHS since the grid size of the parameter space is very large. $\mathcal{X}_{\rm ref}$ includes 189 points corresponding to each day in the time. At each iteration, a new input is acquired from a candidate list $\mathcal{L}_t$ of size 2600 constructed to simplify the optimization process. To generate $\mathcal{L}_t$, 100 parameters are sampled uniformly from the prior each of which is augmented with existing design inputs ($=13 \times 100$) and randomly selected 13 days from the calibration period are paired with another 100 random parameters ($=13 \times 100$). 
\begin{table}
    \centering
    \begin{tabular}{ccc}
        Parameter  & Definition  & Prior \\ \hline
       $\sigma_I$  & rate at which exposed individuals become infectious (inverse) & $\left[2.4, 3.4\right]$ \\
       $\omega_A$  & infectiousness of asymptomatic individual relative to infectious individual & $\left[0.33, 0.99\right]$ \\
       $\gamma_Y$  & recovery rate from symptomatic compartment (inverse) & $\left[3.9, 4.1\right]$ \\
       $\gamma_A$  & recovery rate from asymptomatic compartment (inverse) & $\left[3.9, 4.1\right]$ \\ \hline
    \end{tabular}
    \caption{Epidemiological parameters and their prior ranges.}
    \label{tab:covid_params}
\end{table}

\begin{figure}[ht]
\centering
    \begin{subfigure}{0.43\textwidth}
        \includegraphics[width=1\textwidth]{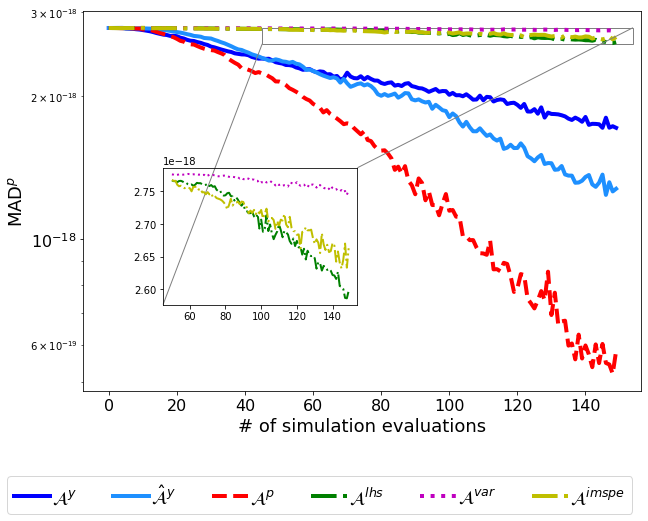}
    \end{subfigure}
    \begin{subfigure}{0.43\textwidth}
        \includegraphics[width=1\textwidth]{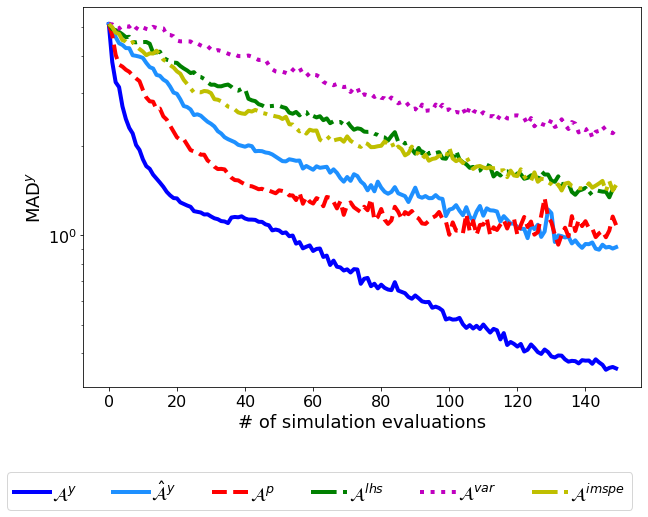}
    \end{subfigure}
    \caption{Comparison of different acquisition functions with the COVID-19 simulation model. The left and right panels compare the accuracy of posterior and field predictions, respectively.}
    \label{covid_pred_res}
\end{figure}

\begin{figure}[ht]
\centering
    \begin{subfigure}{0.485\textwidth}
        \includegraphics[width=1\textwidth]{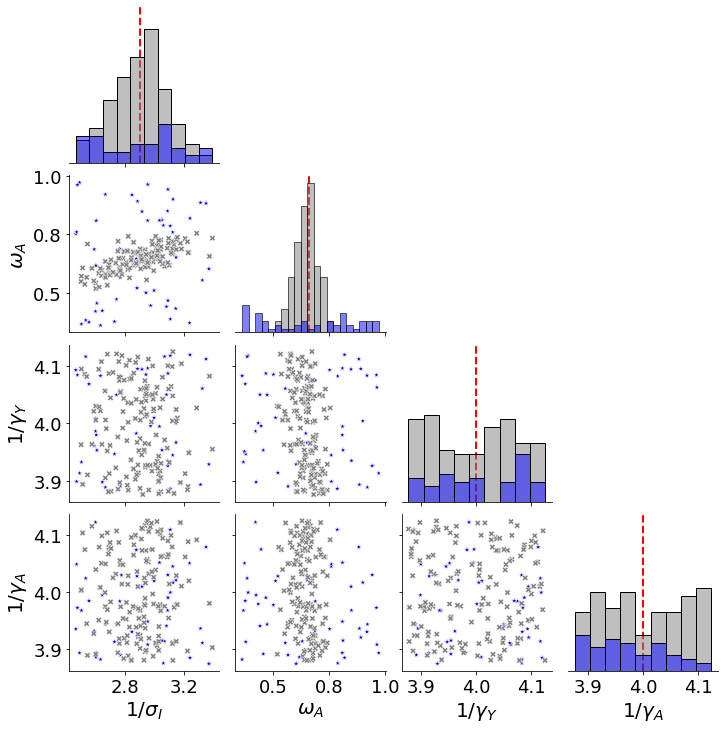}
    \end{subfigure}
    \begin{subfigure}{0.485\textwidth}
        \includegraphics[width=1\textwidth]{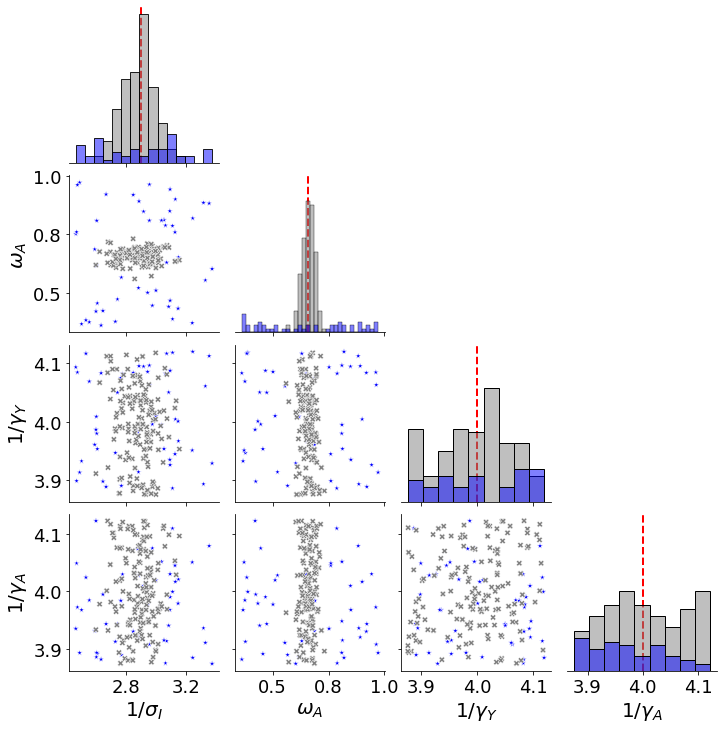}
    \end{subfigure}
    \caption{Pairwise plots for acquired parameters (gray cross markers) via acquisition functions $\mathcal{A}^p$ (left) and $\mathcal{A}^y$ (right). Blue stars are the initial sample obtained randomly from the prior. The red dashed line corresponds to the best-fit parameter $\breve{\thetav}$.}
    \label{covid_param}
\end{figure}

Figure~\ref{covid_pred_res} summarizes the results from 30 replications. As can be seen in the left panel, the acquisition function $\mathcal{A}^p$ outperforms the other functions for predicting the posterior. Even though the posterior predictions improve with \tcb{$\mathcal{A}^{lhs}$, $\mathcal{A}^{var}$, and $\mathcal{A}^{imspe}$} (see the inset zoom), the improvement is negligible as compared to the proposed approaches. The right panel shows that the most accurate field predictions are obtained with $\mathcal{A}^y$. We observe the pairwise scatterplots of acquired parameters for a single replicate of $\mathcal{A}^{p}$ and $\mathcal{A}^{y}$ in Figure~\ref{covid_param}. Additionally, Table~\ref{table:is_covid} provides the average interval score for calibration parameters acquired with each method across 30 replicates. Using the proposed acquisition functions, parameters $1/\sigma_I$ and $\omega_A$ are well-constrained around the best-fit parameters, which are also found as the most influential parameters for calibration in \cite{Surer2021}. Moreover, the acquisition function $\mathcal{A}^{y}$ considers densely the regions around the best-fit parameters, whereas $\mathcal{A}^{p}$ focuses on a wider region since it considers the total uncertainty across the parameter space. Similarly, $\hat{\mathcal{A}}^{y}$ focuses on a wider parameter region of interest since varying $\hat{\thetav}$ values from one iteration to another lead to exploration of the high posterior region. Consequently, while $\hat{\mathcal{A}}^{y}$ achieves lower ${\rm MAD}^p$ values through exploration of the parameter region of interest, $\mathcal{A}^{y}$ obtains the lowest ${\rm MAD}^y$ value through exploitation of the region around $\breve{\thetav}$. \tcb{Additionally, $\mathcal{A}^{imspe}$ performs similarly to $\mathcal{A}^{lhs}$, as both approaches sample from the entire input space. In contrast, $\mathcal{A}^{var}$ shows the poorest performance due to its tendency to sample primarily from the boundaries of the input space.}

\begin{table}[ht]
\footnotesize
    \centering
    \setlength{\tabcolsep}{3pt}
    \begin{tabular}{ccccccc}
          & $\mathcal{A}^y$ & $\hat{\mathcal{A}}^y$ & $\mathcal{A}^p$ & $\mathcal{A}^{lhs}$ & \tcb{$\mathcal{A}^{var}$} & \tcb{$\mathcal{A}^{imspe}$} \\ \hline
       $1/\sigma_I$     & 0.37 & 0.60 & 0.53 & 0.89 & 0.99 & 0.92 \\
       $\omega_A$       & 0.15 & 0.34 & 0.24 & 0.89 & 0.99 & 0.94 \\ 
       $1/\gamma_Y$     & 0.77 & 0.85 & 0.85 & 0.89 & 0.97 & 0.87 \\
       $1/\gamma_A$     & 0.84 & 0.89 & 0.88 & 0.89 & 0.97 & 0.85 \\ \hline
    \end{tabular}
    \caption{The average interval score for each calibration parameter across 30 replicates of each acquisition function.}
    \label{table:is_covid}
\end{table}

\begin{figure}[ht]
\centering
    \begin{subfigure}{0.45\textwidth}
        \includegraphics[width=1\textwidth]{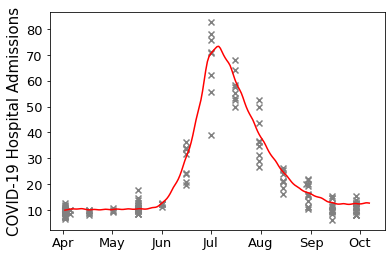}
    \end{subfigure}
    \begin{subfigure}{0.45\textwidth}
        \includegraphics[width=1\textwidth]{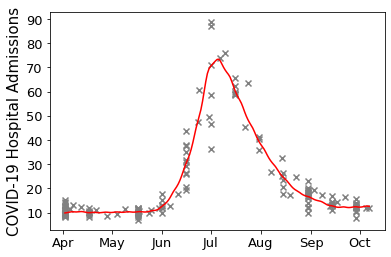}
    \end{subfigure}
    \caption{Simulation outputs (gray cross markers) obtained with the sequential strategy via acquisition functions $\mathcal{A}^p$ (left) and $\mathcal{A}^y$ (right). The red line illustrates the simulation outputs at $\breve{\thetav}$ across design inputs.}
    \label{covid_output}
\end{figure}

Figure~\ref{covid_output} illustrates $n = 150$ simulation outputs collected with $\mathcal{A}^{p}$ and $\mathcal{A}^{y}$ for a single replicate. Both $\mathcal{A}^{p}$ and $\mathcal{A}^{y}$ select simulation outputs concentrated around 13 field data design inputs, leading to improved posterior predictions. The hospital admissions peak around July 2020 in Austin, and the simulation outputs obtained at the time of the peak using LHS range from zero to 500 in the right panel of Figure~\ref{fig:covid_model}. In Figure~\ref{covid_output}, both $\mathcal{A}^{p}$ and $\mathcal{A}^{y}$ select outputs around the region of interest (represented by the red line), and no simulation output is selected from the zero-posterior regions where the peak value is very small (i.e., less than 40 daily admissions) or very large (i.e., larger than 200 daily admissions). Overall, the acquisition function $\mathcal{A}^{p}$ is advantageous when the goal is to better estimate the parameters and understand their relationship through estimating the posterior density. Although $\hat{\mathcal{A}}^{y}$ explores the design space similar to $\mathcal{A}^{y}$, it provides slightly better field predictions than $\mathcal{A}^{p}$ through the end of the procedure since acquiring around 13 field data design inputs is adequate for $\mathcal{A}^{p}$ to successfully predict the field data due to the characteristics of the response surface. On the other hand, $\mathcal{A}^{y}$ takes advantage of targeted and consistent acquisitions around $\breve{\thetav}$ while exploring the design space for improved field predictions without losing too much of the accuracy of posterior predictions.


\section{Conclusion}
\label{sec:conc}

We propose two novel acquisition functions for improved calibration inference in an active learning setting. Our results suggest that exploitation of existing field data design points improves posterior prediction, whereas exploration of the design space along with exploitation improves field predictions. Moreover, the proposed acquisition functions encourage selecting parameters in regions of high posterior density, which is essential to accurately learn both the posterior and field data. This work can be expanded in many directions. One area is to modify the proposed acquisition functions for a field experiment design rather than a simulation experiment design to efficiently infer the calibration parameters. Following that, we further examine both one-shot and active learning settings (see, for example, \cite{Krishna2022} and \cite{Williams2011}) to find new design points to collect the field data for improved calibration of simulation models. Given the cost of performing a real physical experiment, this research would play a transformative role in optimizing the investment by guiding the optimal physical experiment design. In parallel to this, integrating these acquisition functions into the combined field and simulation experiments is the subject of another ongoing work (see comparisons in \cite{Leatherman2017} for the selection of the initial design) since both field and simulation experiments must be carefully designed to effectively use the limited resources. Moreover, sequential design can further benefit from using the proposed acquisition functions at different iterations of the sequential process to adapt to their evolving characteristics in a hybrid way. Alternatively, the performance can be explored with other design criteria such as space-filling designs in conjunction with the proposed acquisition functions. This work implements the sequential procedure in a one-at-a-time fashion. In the case of multiple processors, evaluating the simulation model in parallel with a batch of inputs is computationally more effective than the one-at-a-time procedure. Although our code implementation allows parallel runs, investigating the effect of batch size on the predictive quality and computational savings will be of great interest to practitioners. Similarly, physical experimentalists may prefer conducting a batch of experiments simultaneously due to the complexity of arranging experimental setups, and in such a case, field experiment designs with extensions allowing batch updates would be another line of future development. \tcb{In this work, we utilize stationary GPs to emulate simulation models. However, it is important to note that simulation models often feature non-stationary response surfaces. Addressing non-stationary properties in GPs can vary depending on the specific model employed, necessitating tailored extensions for the proposed acquisition functions. We leave the exploration and development of these extensions for future work.}

\subsection*{Acknowledgement}
The author is grateful for support from the National Science Foundation (NSF) grant OAC 2004601. We gratefully acknowledge the computing resources provided on Bebop, a high-performance computing cluster operated by the Laboratory Computing Resource Center at Argonne National Laboratory.

\bibliographystyle{apalike}
\bibliography{refs}

\appendix

\section{Supplementary Material}

\subsection{Proofs}
\label{app:code}

\subsubsection{Proof of Lemma~3.1}
\label{app:3.1}
        Using Equation~\eqref{eq:posterior}, we obtain $\mathbb{E}[\tilde{p}(\thetav|\y)|\mathcal{D}_{t}] = \mathbb{E}[p(\y|\thetav) p(\thetav)| \mathcal{D}_{t} ] = \mathbb{E}[p(\y|\thetav)| \mathcal{D}_{t}]p(\thetav)$.  Thus, to complete the proof, it is enough to show $\mathbb{E}[p(\y|\thetav)|\mathcal{D}_{t}] = f_\mathcal{N}\left(\y; \, \meanv, \, \Sigmav + \cov\right)$. We drop $\thetav$ from $\model(\thetav)$, $\meanv$, and $\cov$ for the remainder of the proof for brevity.  Using Equations~\eqref{eq:truelike} and ~\eqref{emu_final}, $\mathbb{E}[p(\y|\thetav)|\mathcal{D}_{t}] = {\displaystyle \int} f_\mathcal{N}\left( \y; \, \model, \, \Sigmav \right) f_\mathcal{N}\left(\model; \, \muv_t, \, \Sv_t\right) d \model,$ which is equivalent to
        \begin{align}\label{jointlike_ext}
            \begin{split}
                (2\pi)^{-d}|\Sigmav|^{-1/2} |\Sv_t|^{-1/2} \int \exp\left\{-\frac{1}{2}(\y - \model)^\mathsf{T}\Sigmav^{-1}(\y - \model) -\frac{1}{2}(\model - \muv_t)^\mathsf{T}\Sv_t^{-1}(\model - \muv_t)\right\} d \model.
            \end{split}
        \end{align}
        Equation~\eqref{jointlike_ext} can be expressed in an equivalent form 
        \begin{align}\label{jointlikesubs} 
            (2\pi)^{-d} |\Sigmav \Sv_t|^{-1/2} \int \exp\left\{-\frac{1}{2}(\mathbf{v} + \mathbf{z})^\mathsf{T}\Sigmav^{-1}(\mathbf{v} + \mathbf{z}) -\frac{1}{2}\mathbf{v}^\mathsf{T}\Sv_t^{-1}\mathbf{v}\right\} d \mathbf{v},
        \end{align} 
        where $\mathbf{v} \coloneqq \muv_t - \model$ and $\mathbf{z} \coloneqq \y - \muv_t$.
        Equation~\eqref{jointlikesubs} can be represented in matrix notation as
        \begin{align}\notag
            \begin{split}
                \mathbb{E}[p(\y|\thetav)|\mathcal{D}_{t}] &= (2\pi)^{-d} |\Sigmav \Sv_t|^{-1/2} \int \exp\left\{-\frac{1}{2}\left[{\begin{array}{c} \mathbf{v} \\
                \mathbf{z} \\\end{array}} \right]^\mathsf{T}\left[{\begin{array}{cc} \Sigmav^{-1} + \Sv_t^{-1} & \Sigmav^{-1} \\
                \Sigmav^{-1} & \Sigmav^{-1} \\
                \end{array} } \right] \left[{\begin{array}{c} \mathbf{v} \\
                \mathbf{z} \\ \end{array} } \right]\right\} d \mathbf{v} \\
                &= \int f_\mathcal{N}\left(\left[{\begin{array}{c} \mathbf{v} \\
                \mathbf{z} \\\end{array}} \right]; \,  \mathbf{0}, \, \left[ {\begin{array}{cc} \Sv_t & -\Sv_t \\
                                   -\Sv_t & \Sigmav + \Sv_t \\
                \end{array} } \right]  \right)d \mathbf{v}.
            \end{split}
        \end{align}  
        Following the Gaussian identities in the appendix of \cite[Equation~(A.6)]{Rasmussen2005} completes the proof.

        Similarly, we obtain $\mathbb{V}[\tilde{p}(\thetav|\y)| \mathcal{D}_{t}] = \mathbb{V}[p(\y|\thetav)p(\thetav)| \mathcal{D}_{t}] = \mathbb{V}[p(\y|\thetav)| \mathcal{D}_{t}]p(\thetav)^2$. By definition, $\mathbb{V}[p(\y|\thetav)| \mathcal{D}_{t}] = \mathbb{E}[p(\y|\thetav)^2| \mathcal{D}_{t}] - \mathbb{E}[p(\y|\thetav)| \mathcal{D}_{t}]^2$. From the first proof, we have $\mathbb{E}[p(\y|\thetav)| \mathcal{D}_{t}]^2 = f_\mathcal{N}\left(\y; \, \muv_t, \, \Sigmav + \Sv_t\right)^2$. Thus, it suffices to show that $\mathbb{E}[p(\y|\thetav)^2| \mathcal{D}_{t}] = \frac{1}{2^{d}\pi^{d/2}|\Sigmav|^{1/2}}f_\mathcal{N}\left(\y; \, \muv_t, \, \frac{1}{2}\Sigmav + \Sv_t\right)$. We obtain $\mathbb{E}[p(\y|\thetav)^2| \mathcal{D}_{t}] = {\displaystyle \int} \left(f_\mathcal{N}\left(\y; \, \model, \, \Sigmav\right)\right)^2 f_\mathcal{N}\left(\model; \, \muv_t, \, \Sv_t\right) d \model $, which is equivalent to
        \begin{align}\label{expectedsquared}
            \begin{split}
                &= \frac{1}{(2\pi)^{3d/2}|\Sigmav \Sv_t \Sigmav|^{1/2}} \int \exp\left\{-\frac{1}{2}\left(2(\y - \model)^\mathsf{T}\Sigmav^{-1}(\y - \model) + (\model - \muv_t)^\mathsf{T}\Sv_t^{-1}(\model - \muv_t)\right)\right\} d \model.
            \end{split}
        \end{align}
        Defining again $\mathbf{v} \coloneqq \muv_t - \model$ and $\mathbf{z} \coloneqq \y - \muv_t$, Equation~\eqref{expectedsquared} becomes
        \begin{align}\notag
            \begin{split}
                 &= \frac{1}{(2\pi)^{3d/2} |\Sigmav|^{1/2} 2^{d/2}\left|\frac{1}{2} \Sigmav \Sv_t\right|^{1/2} }\int \exp\left\{-\frac{1}{2}\left[{\begin{array}{c} \mathbf{v} \\
                \mathbf{z} \\\end{array}} \right]^\mathsf{T}\left[{\begin{array}{cc} 2\Sigmav^{-1} + \Sv_t^{-1} & 2\Sigmav^{-1} \\
                2\Sigmav^{-1} & 2\Sigmav^{-1}\\
                \end{array} } \right] \left[{\begin{array}{c} \mathbf{v} \\
                \mathbf{z} \\ \end{array} } \right]\right\} d \mathbf{v} \\
                &= \frac{1}{2^{d}\pi^{d/2} |\Sigmav|^{1/2}}\int f_\mathcal{N}\left(\left[{\begin{array}{c} \mathbf{v} \\
                \mathbf{z} \\\end{array}} \right]; \,  \mathbf{0}, \, \left[ {\begin{array}{cc} \Sv_t & -\Sv_t \\
                                   -\Sv_t & \frac{1}{2}\Sigmav + \Sv_t \\
                \end{array} } \right]  \right) d \mathbf{v}.
            \end{split}
        \end{align} 
        Marginalizing over $\mathbf{v}$ completes the proof.

\subsubsection{Proof of Lemma~3.2}
\label{app:3.2}

For any input $\zb = \left(\xb^\top, \thetav^\top\right)^\top$, recall that $m_{t}(\mathbf{z})$ and $\varsigma^2_{t}(\mathbf{z})$ are the emulator mean and variance at iteration $t$. Suppose that we observe the hypothetical output $\eta^* \coloneqq \eta(\xb^*, \thetav^*)$ for any $\zb^* = \left({\xb^*}^\top, {\thetav^*}^\top\right)^\top$. After seeing the simulation data set $\mathcal{D}_{t+1}$ that includes $\left(\zb^*, \eta^* \right)$ (i.e., $\mathcal{D}_{t+1} = (\zb^*, \eta^*) \cup \mathcal{D}_{t}$), we obtain
    \begin{align}\label{mean_future}
        \begin{split}
            m_{t+1}(\mathbf{z}) &= \left[\kv_t(\mathbf{z})^\top, \, k_t(\mathbf{z}, \mathbf{z}^*)\right] \left[ {\begin{array}{cc} \Kv_t & \kv_t(\mathbf{z}^*) \\
            \kv_t(\mathbf{z}^*)^\top & k_t(\mathbf{z}^*, \mathbf{z}^*) + \upsilon\\
            \end{array} } \right]^{-1} \left[ {\begin{array}{cc} \model_{t} \\
            \eta^* \\
                \end{array} } \right] \\
            &= m_{t}(\mathbf{z}) + \frac{\text{cov}_{t}(\mathbf{z}, \mathbf{z}^*)}{\varsigma^2_{t}(\mathbf{z}^*) + \upsilon} (\eta^* - m_{t}(\mathbf{z}^*)).
        \end{split} 
    \end{align}
Using a similar line of reasoning for both variance and covariance functions, we have
    \begin{align}\label{cov_future}
        \begin{split}
            \varsigma^2_{t+1}(\mathbf{z}) = \varsigma^2_{t}(\mathbf{z}) - \frac{\text{cov}_{t}(\mathbf{z}, \mathbf{z}^*)^2}{\varsigma^2_{t}(\mathbf{z}^*) + \upsilon} \qquad \mbox{and} \qquad
            {\rm cov}_{t+1}(\mathbf{z}, \mathbf{z}') = {\rm cov}_{t}(\mathbf{z}, \mathbf{z}') - \frac{\text{cov}_{t}(\mathbf{z}, \mathbf{z}^*)\text{cov}_{t}(\mathbf{z}', \mathbf{z}^*)}{\varsigma^2_{t}(\mathbf{z}^*) + \upsilon}.
        \end{split} 
    \end{align}
Taking the expected value, variance, and covariance of Equation~\eqref{mean_future}, respectively, provides
    \begin{align}
        \begin{split}
        \mathbb{E}_{\eta^* | \mathcal{D}_{t}}\left[m_{t+1}(\mathbf{z})\right] = m_{t}(\mathbf{z}), \quad 
        \mathbb{V}_{\eta^* | \mathcal{D}_{t}}\left[m_{t+1}(\mathbf{z})\right] = \frac{\text{cov}_{t}(\mathbf{z}, \mathbf{z}^*)^2}{\varsigma^2_{t}(\mathbf{z}^*) + \upsilon}, \quad \mbox{and} \\
        \mathbb{C}_{\eta^* | \mathcal{D}_{t}}\left[m_{t+1}(\mathbf{z}), m_{t+1}(\mathbf{z'})\right] = \frac{\text{cov}_{t}(\mathbf{z}, \mathbf{z}^*)\text{cov}_{t}(\mathbf{z'}, \mathbf{z}^*)}{\varsigma^2_{t}(\mathbf{z}^*) + \upsilon}. \qquad
    \end{split}
\end{align}
Using $\eta^*|\model_{t} \sim \text{N}\left(m_{t}(\mathbf{z}^*), \, \varsigma^2_{t}(\mathbf{z}^*) + \upsilon\right)$ and the transformation in Equation~\eqref{mean_future}, we have
    \begin{align}\label{mean_distr1}
        \begin{split}
            m_{t+1}(\mathbf{z}) | \mathcal{D}_t \sim \text{N}\left(m_{t}(\mathbf{z}), \, \frac{\text{cov}_{t}(\mathbf{z}, \mathbf{z}^*)^2}{\varsigma^2_{t}(\mathbf{z}^*) + \upsilon}\right).
       \end{split} 
    \end{align}
Recall that $\meanv$ represents the mean vector of simulation outputs at field data design inputs paired with $\thetav$ at iteration $t$. Then, Equation~\eqref{mean_distr1} implies 
    \begin{align}\label{mean_distr2}
        \begin{split}
            \muv_{t+1}\left(\thetav\right) | \mathcal{D}_t \sim \text{MVN}(\meanv, \PHI),
       \end{split} 
    \end{align}  
where the $i$th diagonal element of $\PHI$ is $\frac{\text{cov}_{t}\left(\mathbf{z}_i^f, \mathbf{z}^*\right)^2}{\varsigma^2_{t}(\mathbf{z}^*) + \upsilon}$ and $(i,j)$th element of $\PHI$ is $\frac{\text{cov}_{t}\left(\mathbf{z}_i^f, \mathbf{z}^*\right)\text{cov}_{t}\left(\mathbf{z}_j^f, \mathbf{z}^*\right)}{\varsigma^2_{t}(\mathbf{z}^*) + \upsilon}$
with $\mathbf{z}_i^f = \left({\xb^f_i}^\top, \thetav^\top\right)^\top$ for $i, j = 1, \ldots, d$. In addition, Equation~\eqref{cov_future} implies $\mathbf{S}_{t+1}(\thetav) = \cov - \PHI$ and notice that $\mathbf{S}_{t+1}(\thetav)$ does not depend on $\eta^*$. 

For the rest of the proof, we omit $\thetav$ from $\meanv$, $\cov$, $\muv_{t+1}\left(\thetav\right)$, and $\mathbf{S}_{t+1}(\thetav)$ for brevity. From Lemma~\ref{lemma:UQ}, we have $$\mathbb{V}[p(\y|\thetav) | (\zb^*, \eta^*) \cup \mathcal{D}_{t}] = \frac{1}{2^{d}\pi^{d/2}|\Sigmav|^{1/2}} f_\mathcal{N}\left(\y; \, \muv_{t+1}, \, \frac{1}{2}\Sigmav + \Sv_{t+1}\right) - \left(f_\mathcal{N}\left(\y; \, \muv_{t+1}, \, \Sigmav + \Sv_{t+1}\right)\right)^2.$$
Using Equation~\eqref{mean_distr2} and replacing $\mathbf{S}_{t+1}$ with $\mathbf{S}_{t} - \PHI$, we obtain $\mathbb{E}_{\eta^* | \mathcal{D}_{t}} \left( \mathbb{V}[p(\y|\thetav) \left| (\zb^*, \eta^*) \cup \mathcal{D}_{t} \right] \right)$ as
    \begin{align} \label{IEVpart3}
        \begin{split}
            &= \int \frac{1}{2^{d}\pi^{d/2}|\Sigmav|^{1/2}}f_\mathcal{N}\left(\y; \, \muv_{t+1}, \, \frac{1}{2}\Sigmav + \Sv_t - \PHI\right) f_\mathcal{N}\left(\muv_{t+1}; \, \muv_{t}, \, \PHI \right) d\muv_{t+1} \\& - \int \left(f_\mathcal{N}\left(\y; \, \muv_{t+1}, \, \Sigmav + \Sv_t - \PHI \right)\right)^2 f_\mathcal{N}\left(\muv_{t+1}; \, \muv_{t}, \, \PHI \right) d\muv_{t+1}.
        \end{split}
    \end{align}
The rest of the proof follows from \cite{Surer2023}, and we provide the remainder for the sake of completeness. Defining $\mathbf{L} \coloneqq \frac{1}{2}\Sigmav + \Sv_t - \PHI$, $\mathbf{M} \coloneqq  \Sigmav + \Sv_t - \PHI$, and $a_1 \coloneqq \frac{2^{-d}\pi^{-d/2}|\Sigmav|^{-1/2}}{(2\pi)^d|\mathbf{L}\PHI|^{1/2}}$, $a_2 \coloneqq \frac{(2\pi)^{-3d/2}}{|\mathbf{M}\PHI\mathbf{M}|^{1/2}}$, and assuming $\mathbf{L}$ and $\mathbf{M}$ are invertible, \eqref{IEVpart3} is equivalently written as 
    \begin{align} 
        \begin{split} \label{eq:gnew}
        & a_1 \int \exp\left\{-\frac{1}{2} \left(\left(\y - \muv_{t+1}\right)^\top \mathbf{L}^{-1} \left(\y - \muv_{t+1}\right) + \left(\muv_{t+1} - \muv_{t}\right)^\top \PHI^{-1} \left(\muv_{t+1} - \muv_{t}\right) \right)\right\}d\muv_{t+1} \\
        & - a_2 \int \exp\left\{-\frac{1}{2} \left(2\left(\y - \muv_{t+1}\right)^\top \mathbf{M}^{-1} \left(\y - \muv_{t+1}\right) + \left(\muv_{t+1} - \muv_{t}\right)^\top \PHI^{-1} \left(\muv_{t+1} - \muv_{t}\right) \right)\right\}d\muv_{t+1}.
        \end{split}
    \end{align}
    Letting $\mathbf{v} \coloneqq \muv_t - \muv_{t+1}$ and $\mathbf{z} \coloneqq \y - \muv_t$, and writing Equation~\eqref{eq:gnew} in matrix notation yields
    \begin{align}\label{jointlikematrix} 
            \begin{split}
                =& \frac{1}{2^{d}\pi^{d/2}|\Sigmav|^{1/2}} \int f_\mathcal{N}\left(\left[{\begin{array}{c} \mathbf{v} \\
                \mathbf{z} \\\end{array}} \right]; \,  \mathbf{0}, \, \left[ {\begin{array}{cc} \PHI & -\PHI\\
                -\PHI & \mathbf{L} + \PHI\\
                \end{array} } \right]  \right)d \mathbf{v} \\
                &-\frac{1}{2^{d}\pi^{d/2}|\mathbf{M}|^{1/2}} \int f_\mathcal{N}\left(\left[{\begin{array}{c} \mathbf{v} \\
                \mathbf{z} \\\end{array}} \right]; \,  \mathbf{0}, \, \left[ {\begin{array}{cc} \PHI & -\PHI\\
                -\PHI & \frac{1}{2}\mathbf{M} + \PHI\\
                \end{array} } \right]  \right)d \mathbf{v}.
            \end{split}
        \end{align}
    Marginalizing over $\mathbf{v}$ proves that
    \begin{align} \notag 
        \begin{split}
            \mathbb{E}_{\eta^* | \mathcal{D}_{t}} \left( \mathbb{V}[p(\y|\thetav) \left| (\zb^*, \eta^*) \cup \mathcal{D}_{t} \right] \right) = \frac{f_\mathcal{N}\left(\y; \, \muv_{t}, \, \frac{1}{2}\Sigmav + \Sv_t \right)}{2^{d}\pi^{d/2}|\Sigmav|^{1/2}} - \frac{f_\mathcal{N}\left(\y; \, \muv_{t}, \, \frac{1}{2}\left(\Sigmav + \Sv_t + \PHI\right)\right)}{2^{d}\pi^{d/2}|\Sigmav + \Sv_t - \PHI|^{1/2}} .
        \end{split}
    \end{align}

\subsection{Analysis on Initial Sample}
\label{app:initial}

\begin{figure}[ht]
\centering
    \begin{subfigure}{0.4\textwidth}
        \includegraphics[width=1\textwidth]{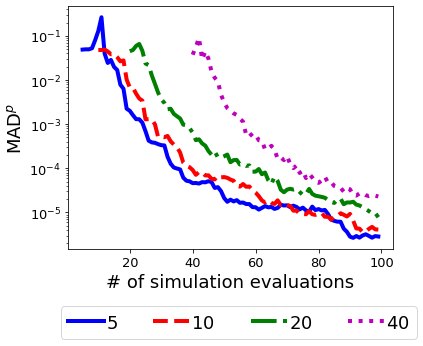}
    \end{subfigure}
    \begin{subfigure}{0.38\textwidth}
        \includegraphics[width=1.05\textwidth]{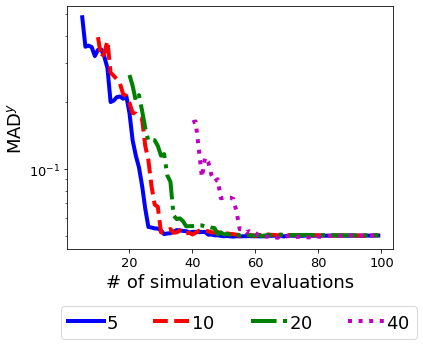}
    \end{subfigure}
    \begin{subfigure}{0.4\textwidth}
        \includegraphics[width=1\textwidth]{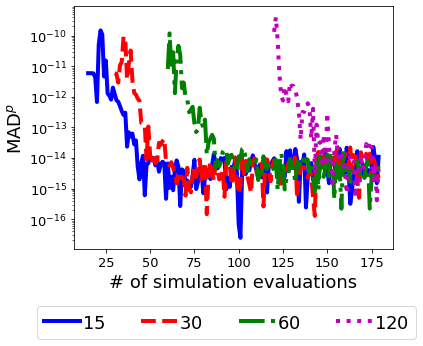}
    \end{subfigure}
    \begin{subfigure}{0.38\textwidth}
        \includegraphics[width=1.05\textwidth]{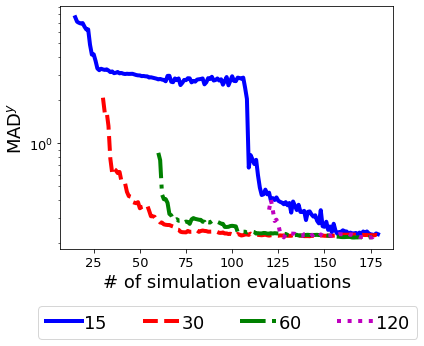}
    \end{subfigure}
    \caption{Experiment results for varying values of initial sample $n_0 \in \{5, 10, 20, 40\}$ ($n_0 \in \{15, 30, 60, 120\}$) using the two-dimensional (three-dimensional) simulation model in Figure~\ref{fig:models_sinf} (Figure~\ref{fig:models_pritam}). The top and bottom panels illustrate results for two- and three-dimensional models, respectively. The left and right panels compare the accuracy of posterior and field predictions using $\mathcal{A}^p$ and $\mathcal{A}^y$, respectively.}
    \label{fig:initial_sample}
\end{figure}

We investigate the effect of the initial sample size $n_0$ on the performance of $\mathcal{A}^p$ and $\mathcal{A}^y$ using the two- and three-dimensional simulation models presented in Section~\ref{sec:synthetic}. We vary the number of observations as $n_0 \in \{5, 10, 20, 40\}$ and $n_0 \in \{15, 30, 60, 120\}$ for the two- and three-dimensional functions, respectively, and summarize the results over a single replicate to illustrate the effect of $n_0$. The algorithm terminates upon reaching a total of 100 simulation evaluations (i.e., $n + n_0 = 100$) for the first example and 180 evaluations (i.e., $n + n_0 = 180$) for the second example. The results are shown in Figure~\ref{fig:initial_sample} for different values of $n_0$. 
 In both examples, $\mathcal{A}^y$ has larger errors for smaller values of $n_0$ early in the algorithm. However, for the two-dimensional model, $\mathcal{A}^y$ with smaller $n_0$ values (i.e., $n_0 = 5$ and $n_0 = 10$) achieves convergence with fewer simulation evaluations compared to those with larger $n_0$ values (i.e., $n_0 = 20$ and $n_0 = 40$) thanks to the fast convergence rate of $\mathcal{A}^y$. In such a case, if the algorithm terminates upon achieving the desired accuracy level, larger $n_0$ values would result in wasted computational resources, especially for simulation models with long run times. On the other hand, for the three-dimensional model, the initial sample size $n_0 = 30$ allows a more thorough exploration of the complex response surface, which in turn improves the overall performance of $\mathcal{A}^y$ as compared to $n_0 = 15$. In this example, exploration with $n_0 = 30$ prevents $\mathcal{A}^y$ from getting stuck on local optimal regions. On the other hand, $\mathcal{A}^p$ takes advantage of its fast convergence rate with smaller initial sample sizes in both examples. Overall, the initial sample size plays an important role in the algorithm's performance and an appropriate initial sample size depends on the complexity of the response surface, computational resources, and convergence rate of the acquisition function for a particular application.

 

\subsection{Details for Experiments with High Dimensional Inputs}
\label{app:highdim}
We test the proposed approaches using varying values for the dimensions of the design and parameter spaces. We maintain the input space dimension at $q + p = 12$ and generate three scenarios similar to those in \cite{Surer2023}. The data generation mechanism for examples with higher dimensional input spaces is provided below.
\begin{itemize}
    \item For the example with $q = 2$ and $p = 10$, we consider $\xb = (x_1, x_2) \in [0, 1]^{2}$, $\thetav = (\theta_1, \ldots, \theta_{10}) \in [-5, 5]^{10}$, and $\eta(\xb, \thetav) = \sqrt{x_1 + x_2} (\theta_1 + \cdots + \theta_{10})^2$. The field data is generated through $y\left(\xb_i^f\right) = \eta\left(\xb_i^f, \thetav = \breve{\thetav}\right) + \epsilon$ with $\epsilon \sim {\rm N}(0, 25)$, $\xb_i^f = \textbf{0.5}_{2}$, $i=1,\ldots,4$, and $\breve{\thetav} = \textbf{0}_{10}$.
    \item For the example with $q = 6$ and $p = 6$, we consider $\xb = (x_1, \ldots, x_6) \in [0, 1]^{6}$, $\thetav = (\theta_1, \ldots, \theta_{6}) \in [-5, 5]^{6}$, and $\eta(\xb, \thetav) = \sqrt{x_1 + \cdots + x_6} (\theta_1 + \cdots + \theta_{6})^2$. The field data is generated through $y\left(\xb_i^f\right) = \eta\left(\xb_i^f, \thetav = \breve{\thetav}\right) + \epsilon$ with $\epsilon \sim {\rm N}(0, 5)$, $\xb_i^f = \textbf{0.5}_{6}$, $i=1,\ldots,4$, and $\breve{\thetav} = \textbf{0}_{6}$.
    \item For the example with $q = 10$ and $p = 2$, we consider $\xb = (x_1, \ldots, x_{10}) \in [0, 1]^{10}$, $\thetav = (\theta_1, \theta_{2}) \in [-5, 5]^{2}$, and $\eta(\xb, \thetav) = \sqrt{x_1 + \cdots + x_{10}} (\theta_1 + \theta_{2})^2$. The field data is generated through $y\left(\xb_i^f\right) = \eta\left(\xb_i^f, \thetav = \breve{\thetav}\right) + \epsilon$ with $\epsilon \sim {\rm N}(0, 1)$, $\xb_i^f = \textbf{0.5}_{10}$, $i=1,\ldots,4$, and $\breve{\thetav} = \textbf{0}_{2}$.
\end{itemize}

In addition to the space-filling design $\mathcal{A}^{lhs}$, we investigate the difference between the proposed acquisition functions $\mathcal{A}^{p}$ and $\mathcal{A}^{y}$ and the two other common acquisition functions using the examples with high dimensional input spaces. As mentioned in the introduction, one common acquisition strategy is to select the next point where the emulator uncertainty is highest \citep{Seo200}. In the experiments, the corresponding method abbreviated by $\mathcal{A}^{var}$ uses
\begin{align} \label{maxvar}
    \begin{split}
        \mathbf{z}^{\rm new} &\in  \argmax_{\zb^* \in \mathcal{L}_t} \varsigma^2_{t}(\zb^*)
    \end{split}
\end{align}
in place of line~6 of Algorithm~\ref{alg:oaat}. One drawback of $\mathcal{A}^{var}$ is that it tends to choose inputs from the boundaries. As an alternative, the integrated mean squared prediction error (IMSPE) considers the emulator uncertainty integrated over the input space to avoid inputs at the boundary locations. The associated method labelled $\mathcal{A}^{imspe}$ replaces line~6 of Algorithm~\ref{alg:oaat} with
\begin{align} \label{imspe}
    \begin{split}
        \mathbf{z}^{\rm new} &\in  \argmin_{\zb^* \in \mathcal{L}_t} \sum_{\zb \in \mathcal{Z}_{\rm ref}} \varsigma^2_{t+1}(\mathbf{z}).
    \end{split}
\end{align}
Here, $\varsigma^2_{t+1}(\mathbf{z})$ is obtained via \eqref{cov_future} and depends on the candidate input $\zb^*$ and $\mathcal{Z}_{\rm ref}$ is a reference set within the $[\mathcal{X}, \Theta]$ space.

At each replication, the initial design of size $n_0$ is randomly selected from a uniform distribution, and all methods $\mathcal{A}^p$, $\mathcal{A}^y$, $\mathcal{A}^{lhs}$, $\mathcal{A}^{var}$, and $\mathcal{A}^{imspe}$ utilize the identical initial sample to ensure a fair comparison. We set $n_0=30$ for the examples with $q=6$, $p=6$ and $q=10$, $p=2$. For the large $p$ case, due to large variability in the performance metrics during the earlier stages of all methods, we set $n_0=50$ when $q=2$, $p=10$. The field data is rerandomized at each replication, and the same field data is employed across different methods within each replication. To construct the discrete set of inputs $\mathcal{L}_t$, first, each unique field data design input is paired with each of 500 parameters sampled from a uniform prior in $\Theta$ space to facilitate the exploitation of field data design inputs. Then, another 1000 inputs are randomly sampled from the prior in $[\mathcal{X}, \Theta]$ space to allow exploration. The reference sets $\Theta_{\rm ref}$, $\mathcal{X}_{\rm ref}$, and $\mathcal{Z}_{\rm ref}$ are constructed with 1500 points generated from LHS. Similar to the experiments in Section~\ref{sec:synthetic}, we compute the performance metrics ${\rm MAD}_t^p$ and ${\rm MAD}_t^y$ at each iteration.

\subsection{Code and Data Availability}
The sequential procedure is implemented in the Python software package Parallel Uncertainty Quantification (PUQ). For practical purposes, the implementation allows users to run a simulation model in a parallel mode as well. PUQ is an open-source software package at \url{{https://github.com/parallelUQ/PUQ/tree/dev/jqt_paper}}. The COVID-19 simulation model is also made publicly available under our repository. The \texttt{README} file contains instructions to install the package and provides a guideline to replicate illustrative examples and a prominent result from the paper.

\end{document}